\documentclass[preprints,article,accept,moreauthors,10pt,a4paper]{mdpi}

\firstpage{1}
\pubvolume{xx}
\issuenum{1}
\articlenumber{1}
\pubyear{2018}
\copyrightyear{2018}
\history{
Published: Tito, E.P.; Pavlov, V.I. Relativistic Motion of Stars near Rotating Black Holes. Galaxies 2018, 6, 61.
 }

\Title{Relativistic Motion of Stars Near Rotating Black Hole}

\Author{Elizabeth P. Tito $^{1}$ 
\orcidA{},
 Vadim I. Pavlov $^{2 *}$ 
\orcidB{}
}

\AuthorNames{Elizabeth P. Tito and Vadim I. Pavlov}

\address{%
$^{1}$ \quad Scientific Advisory Group, Pasadena, CA 91125, USA\\
$^{2}$ \quad Universit\'{e} de Lill\'{e}, Facult\'{e} des Sciences et Technologies, 
F-59000 Lille, France}
\corres{Correspondence: Vadim.Pavlov@univ-lille1.fr}

\abstract{
Formulation of the Lagrangian approach is presented for studying features of motions of stellar bodies with non-zero rest mass in the vicinity of fast-spinning black holes. 
The structure  of Lagrangian is discussed. 
The general method is applied to description of  body motion in the Kerr model of space--time to transition  
to the problem of tidal disruption of elastic bodies by strong gravitational field.  
}

\keyword{Fast--spinning black hole, Kerr space--time, Structure of Lagrangian, Motion regimes 
}

\PACS{04.25.-g; 98.10.+z}  
\begin{document}
\section{Introduction}
\label{se1}

In contrast to the well-known ray-method for describing (in non-flat space-time) motion of both photon and  "test body" -- an idealized conceptualization of a material object with non-zero but small mass (not perturbing space-time around it) -- we formulate a step-by-step Lagrangian approach that not only describes the motion of the test body with non-zero rest mass  (obviously, it must move along a geodesic), but also describes interactions of test--like  bodies in a multi-body system, for example, a system of interacting bodies moving in a given strong gravitational field. 

\begin{figure}[h!]
 \centering
 \includegraphics[width=8cm]{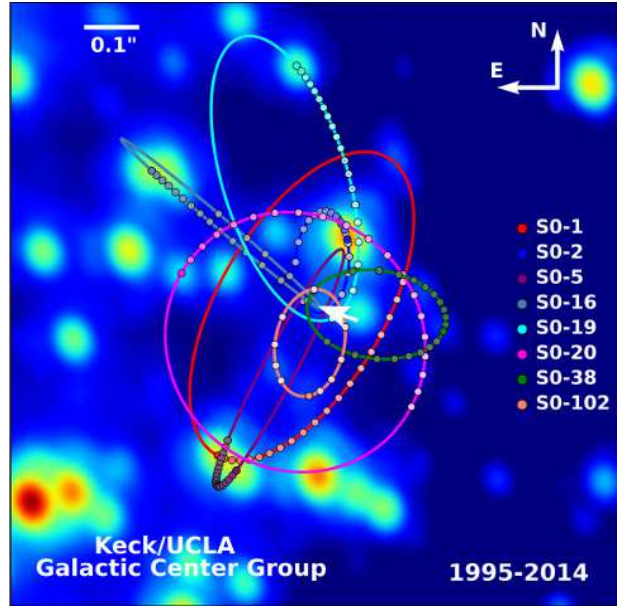}  %\columnwidth
 \caption{
 There exist indications that  the central zone of  our galaxy contains   a compact, possibly rotating, object with mass $\sim 4 \times 10^6 M_\odot$. This super--massive object, thought to be a black hole, exerts strong influence on the dynamics of nearby stars. (See, for example, \citet{g14, r08}, as well as \citet{kt14}.)  The image (from \citep{g14})  shows tracks of the brightest stars near the center of the Milky Way. 
 The image presented here was obtained by UCLA research team based on data obtained by the teams at the W.M.Keck Observatory over the span of two decades. 
 The orbits  plot as dots the star positions at one-year intervals.  The central arrow (added by us) points at the presumed  location of the super--massive black hole. 
}
 \label{bhMW}
 \end{figure}

The need for the ability to conduct detailed examinations of such scenario is obvious.  One vivid example of its importance comes from the astronomical observations 
showing that at the center of our galaxy a super--massive black hole significantly affects the dynamics of nearby stars (see Fig.\ref{bhMW}).   This black hole  is apparently one of the closest such objects to us.  
 It represents an extraordinary,  natural, laboratory  not only for validating  the general relativity theory 
 overall,   
 but also for testing specific theoretical models by comparing them with direct observational data. 
Further improvements in observational techniques combined with continuous (rather than periodic or ad hoc) collection of data can generate invaluable material for fundamental research.

In view of this, 
the importance of the task of formulating the proper procedure for describing motions of bodies, both non-deformable and such that can be torn apart by tidal forces, is apparent.
The framework of the Lagrangian Approach with the relativistically invariant Lagrangian is best-suited for the task. 

The structure of the articles is as follows.
In Sec.~2, we lay out the essentials for working with the space-time metrics near a \emph{rotating}  black hole. This section may be skipped by an experienced reader, 
but it is essential for the completeness of the presentation.
In Sec.~3, we recall the essence of the principle of least action and set up the model of the relativistically invariant Lagrangian for describing motions of bodies near fast-rotating black holes.
Sec.~4 discusses the Euler-Lagrange motion equations for bodies with non-zero mass, that are derived from this principle.
The special case of planar parabolic motion of bodies is considered in Sec.~5.
In Sec.~6, we discuss the obtained expression for Lagrangian in the post-newtonian approximation, which serves as the (complexity-reducing) basis for 
numerical simulations of tidal disruptions of elastic, 
plasma, 
"droplet-like", and so on, bodies, when they enter the close vicinity of a fast-rotating black hole.
Sec.~7 concludes with discussion.

\section{Space--Time Near Rotating Black Hole}
\label{se2}

Practicing any methodology for a long time carries a risk of forgetting the underlying assumptions and choices made during  derivation of core tenets. Therefore, once in a while, the entire framework should be given a fresh review.

\subsection{The Metric:  A Refresher}
\label{sse21}

The geometry of space--time in the vicinity of mass $M$ rotating with angular momentum $J_h$ is described by the Kerr metric (see, for example \cite{mtw73, dfknp97, st04, v08, fz11, tb17}, or any other textbook on the physics of black holes). The Kerr metric describes the gravitational field of a non-charged rotating black hole, and if the radius of the gravitating body is greater that the radius of the event horizon, then in many cases, it describes gravitational fields of other rotating astrophysical objects -- galaxies, stars, and neutron stars.

The interval $ds$ for the Kerr metric in Boyer - Lindquist presentation is found from expression
\begin{eqnarray}
\label{i}
  d\widetilde{s}^2 = \widetilde{g}_{\alpha \beta} dq^{\alpha} dq^{\beta} \rightarrow \nonumber \\
   d \widetilde{s}^2 = (1 - \frac{r_g r}{\Sigma}) c^2 dt^2 +
  \frac{2 r_g r \bar{a}}{\Sigma} \sin^2 \theta c dt d\phi 
   -\frac{\Sigma}{\Delta} dr^2 + \Sigma d\theta^2 - (r^2 + \bar{a}^2 - \frac{r r_g \bar{a}^2}{\Sigma} \sin^2 \theta ) \sin^2 \theta d \phi^2 .
\label{metrics0}
\end{eqnarray}
Further on in this article, we employ metric signature
$diag(+ - - -)$ and  geometric units $G = 1$ and $c = 1$, where
$G$ is the gravitational constant and $c$ is the speed of light in vacuum.
Parameters $q^{\alpha}$ specify the 4-coordinate location of a world point in the space--time, where indices $\alpha$ take values $\alpha = 0,1,2,3$. Coordinates $q^{\alpha}$ are defined from the perspective of the observer located at infinity and characterize any event by a space-time point $q^{\alpha} = (t, r, \theta , \phi)$. Space coordinates $r , \theta , \phi$ are the standard spherical coordinates; this can be easily seen by taking the limit $r \rightarrow \infty$, i.e. $r \gg r_g$ and $r \gg \bar{a}$. In this limit case, the metric is determined by $d\widetilde{s}^2 \rightarrow  c^2 dt^2 - dr^2 - r^2 ( d\theta^2 + \sin^2 \theta d \phi^2 )$. In other words, parameters $(r, \theta , \phi)$ coincide at infinity with the standard spherical coordinates in a flat space-time.

For the region of the space-time where the principle of causality holds, the metric must be $d \widetilde{s}^2 > 0$.

Let us note that $r$ is not the "distance" from the black hole, because in this space-time there exists no central point $r = 0$ in the sense of a particular space-time point (as an event on a valid world-line of a material object).

Coordinates of Boyer - Lindquist merely represent one method of specifying locations of space-time points.

In the proposed Boyer - Lindquist parametrization,  the $z$-axis is co--linear to the angular moment $\textbf{J}_{h}$.
The other parameters in Eq.~(\ref{metrics0}) are:
$r_g$ is the Schwarzschild radius,
$r_g = {2G M} / {c^{2}}$,
$c$ is the light speed,
$G$ is the gravity constant.
Also, for brevity, the following length--scales $\bar{a}$, $\Sigma$, and $\Delta$, have been introduced in Eq.~(\ref{metrics0}):
\begin{eqnarray}
  \bar{a}= \frac{J}{M c}, \quad \Sigma = r^2 + \bar{a}^2 \cos^2 \theta ,
  \quad \Delta = r^2 - r_g r + \bar{a}^2 .
  \label{params0}
\end{eqnarray}
Parameter $\bar{a}$ characterizes the rapidity of rotation of the black hole, $\bar{a} / r_g = J_h c / 2 G M_h^2 $,
however,  calling the parameter the angular velocity of the black hole won't be accurate. In some publications, parameter $\Omega = 2 \bar{a} / r_g$ has been 
used.\footnote{In general, parameter $\bar{a}$ is not small in value for many physically interesting situations.
For example, $J$ of the Sun (assuming uniform rotation) is of order $J \simeq 2 \times 10^{48} \, g \, cm^2 / s$ resulting in $\Omega \simeq 0.185$ -- 
not negligibly small.
}

A key feature of the space--time metric in the vicinity of a rotating black hole is the cross-product term $\sim J_h d t d \phi .$ There exists  \emph{coupling between time and space} that disappears when the black hole's angular momentum goes to zero.\footnote{
To not revisit this topic, let us note that more complicated solutions of Hilbert-Einstein equations exist 
(see for example  \citep{n17}, \citep{ss18} for references).}

In units of length $r_g = 2 G M_h / c^2$ and time $r_g / c$ (when $r = r_g x$, $\bar{a} = r_g a$, and $d\widetilde{s}^2$ is replaced with $r_g^2 ds^2$), the square of dimensionless interval in the vicinity of the rotating black hole in the Kerr metric can be written as
\begin{eqnarray}
  \label{i1}
  ds^2 = g_{\alpha \beta} dq^{\alpha} dq^{\beta} = \nonumber \\
  \bigg[  (1 - \frac{x}{\zeta^2}) dt^2 +
  2 a \frac{x}{\zeta^2} \sin^2 \theta d\phi dt 
  - \frac{\zeta^2}{\chi} dx^2 - \zeta^2 d\theta^2 - \frac{\Lambda}{\zeta^2} \sin^2 \theta d \phi^2 \bigg],
  \label{metrics}
\end{eqnarray}
where 
\begin{eqnarray}
  \zeta^2 = x^2 + a^2 \cos^2 \theta , \quad \chi = x^2 - x + a^2 , 
  \Lambda = 
   ( x^2 + a^2 )^2 - (x^2 - x + a^2) a^2 \sin^2 \theta.
  \label{params}
\end{eqnarray}
Setting $a = 0$  in Eqs.~(\ref{metrics}) and (\ref{params}) gives the Schwarzschild metric.\footnote{
From the definition of interval $ds^2$, it follows that when $x = Const \rightarrow \infty,\, \theta = Const$ and $\phi = Const$ (i.e., the non-moving observer is at the infinity),
the interval $d s^2 = d t_{(\infty]}^2 > 0$. For a clock located at the point $x = Const \neq \infty,\, \theta = Const$ and $\phi = Const$, the same interval is $d s^2 = (1 - \frac{x}{\zeta^2}) dt_{(x)}^2 )$. It follows then that
$\Delta t_{(x)} / \Delta t_{(\infty]}   = (1 - x / {\zeta^2})^{-1/2} > 1$. This simply shows the well--known red--shift for a clock in the gravitational field compared with a clock at infinity where the gravitational field is absent, is
$\Delta t_{(\infty]} < \Delta t_{(x)}$.
}

\begin{figure}[h!]
    \centering
    \includegraphics[width=7cm]{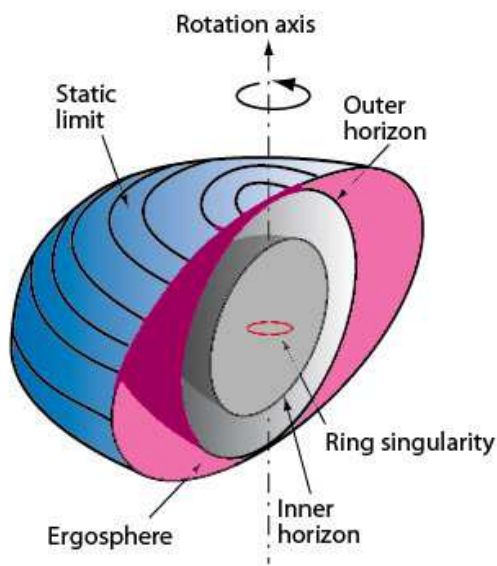}                   %\columnwidth
    \caption{
    Schematic depiction of the black hole structure.
    }
    \label{kerr-bh}
\end{figure}

Eqs.~(\ref{metrics}) and (\ref{params}) show that the Kerr metric has two physically relevant surfaces in the chosen coordinate representation (Fig.~\ref{kerr-bh}).
(In fact, there are four of them, but the ones "beneath" the event horizon are not considered here.)

The \emph{inner} surface corresponds to an \emph{event horizon} similar to the one observed in the Schwarzschild metric -- this occurs where the purely radial component of the metric goes to infinity. Solving equation $g_{11}^{-1} = 0$ yields two solutions, one of which is $x_H = (1 +  {\sqrt {1 - 4 a^{2}}} )/2 \leq 1$.\footnote{
Let us remind that the location of the event horizon $r_{H}$ is determined by the larger of the two roots of $g_{11}^{-1} =0.$ When $\bar{a} > r_g /2 $ (i.e., in the usual units, $J c > G M^2$), there are no real-value solutions to this equation, and there is no event horizon. With no event horizons to hide it from the rest of the universe, the black hole ceases to be a black hole and will instead be a "naked" singularity.
}
Another singularity occurs where the purely temporal component of the metric tensor changes sign from positive to negative.
Again, solving equation $g_{00}=0$ (which has two solutions) yields solution $x_{E} = (1 + {\sqrt {1- 4 a^{2}\cos ^{2}\theta }})/2 $. Due to the presence of the term with $cos^2 \theta$ underneath the radical sign, the outer surface (static limit) resembles a flattened sphere that touches the inner surface at the poles of the rotation axis, where the latitude $\theta$ equals $0$ or $\pi$.

The domain between these two hyper-surfaces is called the \emph{ergosphere}. Within this domain, the purely temporal component of metric tensor is negative, i.e., the time acts as a purely \emph{spatial} coordinate.

The main property of the ergosphere is that, in view of $g_{00} < 0,$ no object can remain at rest with respect to the frame of reference of the remote observer (see \citep{ll69}), because when $x, \theta, \phi = Const$,  Eq.~(\ref{i1}) yields $d s^2 < 0$. In other words, in this case, the interval is not time-like, as it should be for a world-line of a material point-body.
Variable  $q^0 = t$, therefore, loses its time-like property. In the ergosphere, an object cannot possess
$\phi = Const$, because then the cross-term turns to zero since $g_{t \phi} \equiv g_{03}$,  and no circumstances can produce
$d s^2 > 0$ and thus assure the causal linkage along the world-line. Therefore, a particle must necessarily rotate around the axis of symmetry of the gravitational field.
Wherein, changing of radial coordinate $x$ is possible, i.e., $d x \neq 0$. This means that particles may penetrate the ergosphere from the outside and may exit into the outer space. Naturally, for the remote observer, the duration of the process of "falling onto" the static limit may seem infinitely long, as if "frozen". With respect to his own clock, the observer in the vicinity of the black hole crosses the boundaries in a finite time.

In view of the unavoidable co-rotation of bodies  in the ergosphere, it is natural to rewrite the metric $d s^2$ in this domain as follows:
\begin{eqnarray}
  d s^2 = g_{00} ( d t + \frac{g_{03}}{g_{00}} d \phi )^2 + g_{11} (d r)^2 + g_{22} (d \theta)^2 - 
  \bigg( \frac{g_{03}^2 - g_{\phi \phi} g_{00} }{g_{00}} \bigg) (d \phi)^2 ,
  \label{me-erg}
\end{eqnarray}

This expression clearly reveals that time--coordinate $\xi$, determined by the relationship $d \xi = d t + ({g_{03}} / {g_{00}}) d \phi$, behaves in the ergosphere where $g_{00} < 0$, as a space-coordinate, while coordinate $\phi$ starts playing the role of the time-coordinate.

This "switching" of the notions/concepts -- (time-$t$, space-$\phi$) 
being replaced by 
(time-$\phi$, space-$t$) -- 
indicates (in full agreement with the spirit of the principle of relativity) that a "rigid coordinate grid"
(for example, Boyer - Lindquist parametrization)
\emph{cannot} span everywhere, from the infinity and into the ergosphere.  The meaning of parameters defining the world-point for a body, can change when transitioning from one domain of space-time into the other.

Furthermore, singularities of both, the event horizon $S_H$ and the static limit $S_E$, are illusory.
They become singularities only within the "unfortunate" choice of coordinates. Just like Boyer - Lindquist coordinates conveniently turn into normal spherical coordinates at the infinity (far away from the black hole), the space-time near the black hole can be smoothly described (without singularities) by an appropriate choice of coordinates (see, for example, Kruskal--Szekeres, or Lemaitre, 
or Eddington-Finkelstein, 
coordinates).\footnote{
This is already evident from the fact that the determinant of $g_{\alpha \beta}$,  $g = - r^4 \sin^2 \theta \neq 0$,  has no singularity at $r \rightarrow r_g$. Also, making a transformation of the 4-coordinates $(r, t) \rightarrow (R, \tau)$  of the form $c t =\pm c \tau \pm \int d r  (r_g / r)^{1/2} (1 - r_g / r)^{-1}$ and $R = c \tau + \int d r  (1 - r_g / r)^{-1}  (r_g / r)^{-1/2}$, for which $R - c \tau = (2/3) r^{3/2} / r_g^{1/2}$, the element of interval at the Schwarzschild metric becomes 
$d s^2 = c^2 d\tau^2 - [ (3/2 r_g)(R - c \tau) ]^{-2/3} dR^2 -  [ (3/2)(R - c \tau) ]^{4/3} (d \theta^2 + \sin^2 \theta d \phi^2)$. 
Thus, in the coordinates $(R, \tau)$, the singularity at the Schwarzschild surface, where $ (3/2)(R - c \tau) = r_g$,  is absent, but the metric is nonstationary (see \citet{ll00}, \S 102).
}

Pathological appearance of singularities in "poorly"-chosen systems of coordinates can be easily understood by considering the spherical coordinate system when $\theta = 0$ (think of the North Pole on the globe). For $\theta = 0,$  the contribution to the interval $d s^2$ from the term $r^2 \sin^2 \theta d \phi^2$  always equals zero -- \emph{any} value of $\phi$ describes the same world-point. The $\phi$-coordinate appears to "fold" thus creating an \emph{illusory singularity}, which is clearly a product of the "unfortunate" choice of coordinates, and which "disappears" in another coordinate system.

Finally, for a black hole, $a \leq 1/2$ always.   The rapidity of rotation has its extremum at $a = 1/2$.
A quickly rotating object (QRO) with $\Omega = 2 a =  1$ is a maximally rotating QRO. 
Some numerical studies, see for example \cite{dfknp97}, 
have modeled the effects even with $\Omega \rightarrow 1$.

\subsection{The Carter Solution}
\label{sse22}

The characteristic geodesics of the Kerr black-hole space--time have been extensively studied since the pioneering work of Carter \citep{c68}.
See also references and discussions in \citep{ll69}, \citep{bpt72}, \citep{st83}, \citep{d86}, \citep{ch98}, \citep{t03}, \citep{cmbwz09}, \citep{hhls10}, \citep{w12}, \citep{h12}, \citep{ss17}.
Carter suggested deriving the motion equations for test-particles in the Kerr metric based on the obvious condition for covariant 4-velocity $u_\alpha$
\begin{eqnarray}
  \label{c0}
 g^{\alpha \beta} u_{\alpha} u_{\beta} = 1.
\end{eqnarray}
Index raising and lowering occur in accordance with standard rules.

After defining $m u_{\beta}$ as linear momentum, and thus introducing the relativistically-invariant "Lagrangian"
\begin{eqnarray}
  \label{cÄ}
 L = \frac{m}{2} g^{\alpha \beta} u_{\alpha} u_{\beta} \, ,
\end{eqnarray}
Carter proposed introducing the scalar function $S$ ("action") to obtain (with
$m u_{\beta} \equiv \partial L / \partial u^\beta = \partial S / \partial x^{\beta}$) the equations similar  to the
Hamilton--Jacoby equations in mechanics (and eikonal in optics) in the form:
\begin{eqnarray}
  \label{c1}
 g^{\alpha \beta} \frac{\partial S}{\partial x^{\alpha}} \frac{\partial S}{\partial x^{\beta}}=  m^2 .
\end{eqnarray}

Since $g^{\alpha \beta}$ do not explicitly depend on $t$ and $\phi$, the scalar function $S$ (here all variables, as well as coordinates $r$ and $t$, are measured in units $[G] = [M] = [c] = 1$) can be written in the form
\begin{eqnarray}
  \label{c2}
 S = - e t + l_\phi \phi + S_r (r) + S_\theta (\theta ),
\end{eqnarray}
where $e$ is interpreted as the energy of the test--particle, and $l_\phi$ is the projection of the angular momentum onto the axis of rotation. After substitution and redefinition of the time-coordinate, quantities dependent only on $\theta$ and only on $r$ separate. Subsequently, the resulting equations of motion (which we do not list here, but which can be found in any review or textbook on the subject, such as
\citep{bpt72, d86, ch98, cmbwz09, hhls10, ss17}),
represent the first integrals of the equations for geodesic lines. These equations contain $m$, the rest--mass of the particle. For photon, $m = 0.$
In addition to the conservable quantities $e$ and $l_\phi$, Carter also noted the existence of one more conservable quantify -- the separation constant $Q$, whose meaning can be grasped by considering limit $r \rightarrow \infty$ \citep{d86}
\begin{eqnarray}
Q  \rightarrow l^2 - l_\phi^2 - a^2 ( E^2 — m^2 ) \cos^2 \theta_{\infty}.
\end{eqnarray}
Here,  $L^2$ is the constant defined as the square of the full initial momentum.  Obviously, in the case of non-rotating body, when $a = 0$ and the Kerr metric transitions into the Schwarzschild metric, quantity $Q$ is proportionate to the square of the full conservable angular momentum of the test-object.  Furthermore, integral $Q$ is 
analogous to 
the Laplace vector in the Kepler problem of the Newtonian theory of gravity.

Time--like  geodesics  in  this  geometry  admit  four  conservable  quantities:  energy  per unit mass of the particle $e$, angular momentum per unit mass $l_z$, Carter constant of separation\footnote{
In
\cite{h12}, 
the authors, 
citing \citep{bpt72}, present another expression:
 $ Q = p_\theta^2 + \cos^2 \theta [- a^2 p_t^2 + p_\phi^2 / \sin^2 \theta ]$, with $p_t = - E $ total energy and $p_\phi = L_z $ component of angular momentum parallel to the symmetry axis.
}
 $C$, and the norm of the four velocity $u^{\alpha} = dq^{\alpha} / d\tau $, given by
\begin{eqnarray}
e = - u_0 = - g_{00} u^0 - g_{0 \phi} u^{\phi} , \label{4a} \\
l_z = u_{\phi} = g_{\phi 0} u^0 + g_{\phi \phi} u^\phi ,\label{4b} \\
C = \rho^4 (u^\theta)^2 + \sin^{-2}\theta \, l^2_z + a^2 \cos^2 \theta (1-E^2) , \label{4c} \\
g_{\alpha \beta} u^{\alpha \beta} u^{\beta} = - 1. \label{4d}
\end{eqnarray}
The version of the Carter constant used 
in 
 \citep{w12} has the property that, in the Schwarzschild limit ($a \sim J_h \rightarrow 0$), the constant $C \rightarrow l^2$, where $l$ is the total conservable angular momentum of the test body. Solving Eqs.~(\ref{4a}) and (\ref{4b}) for $u^0$ and
$u^\phi$, and Eq.~(\ref{4c}) for $u^\theta$, one obtains
\begin{eqnarray}
u^0 = \frac{E g_{\phi \phi} + l_z g_{0 \phi}}{\Delta \sin^2 \theta} , \quad
u^\phi = - \frac{E g_{0 \phi} + l_z g_{0 0}}{\Delta \sin^2 \theta} , \\
u^\theta = \rho^{-2}
\bigg[C - sin^{-2} \theta l^2_z - a^2 \cos^2 \theta (1 - E^2) \bigg]^{1/2}.
\label{5}
\end{eqnarray}
Substituting these expressions into Eq.~(\ref{4d}) and solving it for $u^r$, yields, after some manipulations, 
\begin{eqnarray}
( u^r )^2 = \frac{r^4}{\rho^4} \bigg( (1 + \frac{a^2}{r^2} + \frac{2 M a^2}{r^3}) E^2
- \frac{\Delta}{r^2 (1 + C r^{-2})} + 
\frac{a^2 l^2_z}{r^4} - \frac{4 M a E l_z}{r^3}\bigg) .
\end{eqnarray}
This equation is used to study the turning points of the radial motion.

\section{Action, Lagrangian and Equations of Motion}
\label{se3}

Routine use of geodesic equations in numerical simulations may produce a habit so ingrained that the technique's theoretical foundation can become misremembered.

Indeed, just to refresh readers' memory, the fact that non-zero-mass test particles move along geodesic lines is a \emph{consequence} of the Principle of Least Action, not its cause. 
The motion of test particles in the given gravitational field results from, and must be found from, extremum for \emph{Action} $S$, not the other way around. 
Any thinking that bodies' movement along geodesic lines has to lead to an extremum of something, is an erroneous confusion of the theoretical basics.

Consequently, the goal of the subsequent discussion is to structure a consistent technique for calculating  trajectories of interacting bodies, that employs  the "correct"~  relativistically-invariant Lagrangian.

\subsection{Principle of Least Action}
\label{sse31}

For a conservative system, action is defined as an integral of Lagrangian $L$ (which is a function of 4-coordinates and 4-velocities)  over some affine
parameter $\eta$,  for example, over the proper time $\tau$.

The choice of Lagrangian is the key point in the construction of the theory. 
This choice must be grounded in common sense, and its validity must be confirmed by experimental observations.

The law of motion for any body, including the test particle, depends on the form of Lagrangian.  From this perspective, the trajectory of the body is model-dependent. 
Some works (besides the above-mentioned work of Carter, see, for example, \citet{r15, ss17} and bibliography cited therein) have offered to use as a Lagrangian, expression
\begin{eqnarray}
  \label{i3c}
  L = \frac{m}{2} g_{\alpha \beta} q_{, \tau}^{\alpha} q_{, \tau}^{\beta} \equiv \frac{m}{2} s_{, \tau}^2
\end{eqnarray}
where, as common, symbol $f_{, \tau}$ denotes the derivative of function $f$ with respect to the proper time $\tau$.
For example, \citet{ss17} stated that "the motion of a neutral particle can be illustrated by the Lagrangian"  Eq.~(\ref{i3c}).
However, such liberal choice of Lagrangian isn't substantiated by physical considerations,
but is based on the resemblance with 
the Lagrangian of a free particle in the Newtonian mechanics, where the usual velocity is substituted by the 4-velocity to assure relativistic invariance.

We posit that for a Lagrangian a "square root form" is more substantiated.

Indeed, following \cite{ll69}, the action $S$ of a test particle must depend only on the interval between two world-points, i.e. $S \sim \int ds$ where  $d s$ is the "distance" between two infinitely close neighboring world-points.
Having in mind Eq.~(\ref{i1}), we can rewrite the action as
\begin{eqnarray}
\label{s0}
S = - m \int d \eta \frac{d s}{d \eta} = - m \int d \eta \sqrt{ g_{\alpha \beta} q_{, \eta}^{\alpha} q_{, \eta}^{\beta} }
\end{eqnarray}
where $\eta$ is some scalar parameter, and where we added some dimensional factor $m$, the essence of which is determined by making a limit transition to 
the flat space--time and the newtonian classical case. 
The principal reason for choosing the action in the form of Eq.~(\ref{s0}) is
that it is invariant when changing time--scales or when rescaling $\eta \rightarrow \eta
(\eta ')$, i.e. $\eta$ is an affine parameter and can be chosen arbitrary. For example, we can introduce
(dimensionless) parameter $\tau$:  $d s^2 = r_g^2 d \tau^2$. The meaning of $\tau$ is that it
represents the (dimensionless) time measured by a non-moving oberver (located at some fixed spatial coordinates in a
moving frame)  in a frame where gravitational field determined by $g_{00}$ is locally compensated, i.e. in the frame which
travels along the geodesic. In other words, the observer follows a time--like world-line, i.e., it is "falling freely".
From Eq.~(\ref{i1}), we obtain that in this case
\begin{eqnarray}
  \label{i2}
  g_{\alpha \beta} q_{, \tau}^{\alpha} q_{, \tau}^{\beta} =  1 \quad \rightarrow \quad  q_{\alpha, \tau} q_{, \tau}^{\alpha} =  1.
\end{eqnarray}

The integrand in Eq.~(\ref{s0}) multiplied by a constant, giving $- m \sqrt{g_{\alpha \beta} q_{, \eta}^{\alpha} q_{, \eta}^{\beta} }$, is the Lagrangian $L$. It is a scalar which depends on 4-coordinates $q^{\alpha}$ and 4-velocities $q_{, s}^{\alpha} = u^{\alpha}$.

\subsection{The Equivalence Principle}
\label{sse32}

According to the special theory of relativity, the action of a free particle moving in a flat Minkowski space-time, has the following form:
\begin{eqnarray}
  \label{ä1}
  S = - m \int d s \sqrt{ \eta_{\alpha \beta} x_{, s}^{\alpha} x_{, s}^{\beta} } .
\end{eqnarray}
Here $x^{\alpha} = (t, x^k)$ are Cartesian coordinates,
$\eta_{\alpha \beta}$ is Minkowski diagonal tensor with signature $(+,- - -)$, $s$
is the affine parameter, which may be, for example, interval $s$  or proper time  $\tau$ (here $d s^2 = c^2 d \tau^2$).

When transitioning to curvelinear coordinates or into a non-inertial system of coordinates,
$x^{\alpha} \rightarrow q^{\alpha}$, the action takes form
\begin{eqnarray}
  \label{ä2}
  S = - m \int d \eta \sqrt{g_{\alpha \beta} q_{, \eta}^{\alpha} q_{, \eta}^{\beta} } ,
\end{eqnarray}
where
\begin{eqnarray}
\label{ä3}
g_{\alpha \beta}= \eta_{\mu \nu} \frac{\partial x^{\mu}}{\partial q^{\alpha}} \frac{\partial x^{\mu}}{\partial q^{\alpha}}  .
\end{eqnarray}

Trajectories corresponding to the extremum of the action, are called geodesics for metric $g_{\alpha \beta}$.

According to the principle of equivalence, gravity must act in exactly the same way onto a test particle with mass $m$.
However, in contrast with the previously discussed case in which the reference system is inertial, the metric of the non-flat space-time cannot be presumed to converge globally as $g_{\alpha \beta} \rightarrow \eta_{\mu \nu}$. It is the presence of the space-time curvature that distinguishes gravitational forces from the inertia forces.

\section{Euler--Lagrange Equations of Motion}
\label{se4}

\subsection{Extremum of Action}
\label{sse41}
In view of the discussion above, the Euler--Lagrange equations of motion follow from the Principle of Least Action\footnote{
Geometrically, geodesics are the curves of extremal length between two points in space-time. The geodesics of Minkowski space-time governed by the special theory of relativity, are 4--dimensional straight lines. The case with $ds^2 > 0$ describes time--like geodesics, the world--lines of free material particles. Photons - massless particles - travel at the speed of light along the null geodesics ($ds^2 = 0$). Within the general relativity theory, free particles travel along geodesics of the curved space--time.
}
\begin{eqnarray}
  \label{i3}
  \delta S = - m \int d\eta \delta \, \sqrt{g_{\alpha \beta} q_{, \eta}^{\alpha} q_{, \eta}^{\beta} } 
  = 0, \quad   i.e.   \quad  - \frac{m}{2} \int d\eta \frac{1}{ \sqrt{ g_{\alpha \beta} q_{, \eta}^{\alpha} q_{, \eta}^{\beta}} }
  \delta ( g_{\alpha \beta} q_{, \eta}^{\alpha} q_{, \eta}^{\beta} ) = 0.
\end{eqnarray}

Having in mind Eq.~(\ref{i2}) on the geodesic line, we find that equation $\delta S = 0$ is reduced to equation $\delta ( g_{\alpha \beta} q_{, \tau}^{\alpha} q_{, \tau}^{\beta} )  = 0$, which permits formal finding of geodesics from
"Lagrangian" (\ref{i3c}).\footnote{
For example, in \citep{c18} (p.106) manipulation analogous to  Eq.~(\ref{i3})  was used for a radical-containing functional.  
}

 Lagrangian~(\ref{i3c}) may be used instead of the "correct" Lagrangian
\begin{eqnarray}
  \label{i3d}
  L = - m \sqrt{g_{\alpha \beta} q_{, \eta}^{\alpha} q_{, \eta}^{\beta}}
\end{eqnarray}
\emph{only} for finding geodesic lines, but Eq.~(\ref{i3c}) does not yield the correct expression for the particle's full energy.

The choice of Lagrangian in form~(\ref{i3d}) is natural because it produces affine invariance and correct limit for the particle's full energy in the case when the field is absent.
Indeed, in limits $x \rightarrow \infty$, i.e. $x \gg 1$, the Lagrangian - the explicit form of which 
for the Kerr model is
\begin{eqnarray}
  \label{i6}
L = - m \bigg( (1 - \frac{x}{\zeta^2}) t_{, \tau}^2 +
  2 a \frac{x}{\zeta^2} \sin^2 \theta \phi_{,\tau} t_{,\tau} 
  - \frac{\zeta^2}{\chi} x_{, \tau}^2 - \zeta^2 \theta_{, \tau}^2 -
  \frac{\Lambda}{\zeta^2} \sin^2 \theta \phi_{, \tau}^2 \bigg)^{1/2} ,
\end{eqnarray}
and where for brevity $\zeta^2, \chi, \Lambda$ are defined by expressions~(\ref{params}) -- becomes
\begin{eqnarray}
L = - m \sqrt{t_{, \tau}^2 - ( x_{, \tau}^2 +  x^2 \theta_{, \tau}^2 + x^2 \sin^2 \theta \phi_{, \tau}^2)} + ...
\end{eqnarray}

Selecting  time $t$ measured by the clock of a remote non-moving observer as the affine parameter $\tau$,
we obtain $t_{, \tau}=1$,
and since now $ x_{, t}^2 +  x^2 \theta_{, t}^2 + x^2 \sin^2 \theta \phi_{, t}^2  = v^2$ -- 
the square of 3-velocity (in the spherical coordinates) of a particle that travels relative to the remote stationary observer -- we obtain the correct expression for Lagrangian of a relativistic free particle (see \cite{ll00}) $L = - m \sqrt{1 - v^2}$ as seen by the remote observer located in the reference frame relative to which the particle is moving.
This follows from $d \tau L = inv$.

Eq.~(\ref{i3d}) also yields the correct expression for the energy of the relativistic particle.

\subsection{Equations of Motion for Test Particle}
\label{sse42}

The equations of motions are
\begin{eqnarray}
  \label{i8}
\delta S = 0 \quad \rightarrow \quad \frac{d}{d\tau} \frac{\partial L}{\partial q_{,\tau}^{\alpha}} -
\frac{\partial L}{\partial q^{\alpha}} = 0 .
\end{eqnarray}
After substitution of general  Eq.~(\ref{i3d}) in Eqs.~(\ref{i8}) and direct calculations of derivatives, 
%For the chosen Lagrangian, 
these equations take  form
\begin{eqnarray}
  \label{i4}
  g_{\alpha \beta} q_{, \tau \tau}^{\beta} + \Gamma_{\alpha \beta \gamma} q^{\beta}_{, \tau} q^{\gamma}_{,
  \tau} = 0 ,
\end{eqnarray}
which obviously must coincide, and does indeed coincide, with geodesic equation for a test particle 
 trajectory.
It is also obvious that by multiplying  the inverse matrix  of the metric tensor $g_{\mu \nu}$, we can obtain
\begin{eqnarray}
  \label{i5}
  q_{, \tau \tau}^{\alpha} + \Gamma^{\alpha}_{\beta \gamma} q^{\beta}_{, \tau} q^{\gamma}_{,
  \tau} = 0 .
\end{eqnarray}
The Christoffel symbols are given by $\Gamma_{\alpha \beta \gamma} = \frac{1}{2} (g_{\alpha \beta ,
\gamma} + g_{\alpha \gamma, \beta} - g_{\gamma \beta , \alpha})$ and $\Gamma^{\alpha}_{\beta \gamma} =
g^{\alpha \mu} \Gamma_{\mu \beta \gamma}$ (see any textbook on relativistic physics). 

Let us note that the connection between symbols $\Gamma^{\alpha}_{\beta \gamma}$
and metric tensor $g_{\alpha \beta}$
arose as a \emph{consequence} of the Principle of Least Action with square root Lagrangian Eq.~(\ref{s0}).
Since derivative $q_{, \tau \tau}^{\alpha}$ is the generalized 4-acceleration, then it is natural to call quantity
$- m \Gamma^{\alpha}_{\beta \gamma} q^{\beta}_{, \tau} q^{\gamma}_{,\tau}$ as 4-force, and 
$\Gamma^{\alpha}_{\beta \gamma} q^{\beta}_{, \tau} q^{\gamma}_{,\tau}$
as gravitational field strength 
(\citet{ll00}, \S\S 85, 87 and \citet{w72}, \S 1, chapter 4).

\subsection{Hidden Symmetry}
\label{sse43}

The above--written machinery is useful because it readily produces the laws of conservation: if Lagrangian $L$ is independent of a generalized coordinate  $q^{\nu}$ (the ignorable coordinate) then quantity $p_{\nu} = \partial L / \partial q_{,\tau}^{\nu}$ is a constant of motion. For Eq.~(\ref{i6}) with Lagrangian $L (x, x_{, \tau}, \theta, \theta_{, \tau}, \underline{\phi}, \phi_{, \tau}, \underline{t}, t_{, \tau})$, the three ignorable coordinates are $\phi $, $t$ and proper time $\tau$, thus producing laws whose physical meanings are the conservation of momentum $p_{\phi} = \partial L / \partial \phi_{,\tau}$, the conservation of full energy $E = - \partial L / \partial t_{,\tau} $, and the conservation of invariant
$Q = q_{, \tau}^{\nu} \partial_{q_{, \tau}^{\nu}} L - L$ where $\nu = 0,1,2,3$,
which for Lagrangian~(\ref{i6}) is identical to zero. Finally, the set should also include  the invariant Eq.~(\ref{i2}).

\subsection{Trajectory}
\label{sse44}

The trajectory of the test particle can be found in a straightforward way by using the hidden symmetry of the Lagrangian 
 for the Kerr model   Eq.~(\ref{i6})
\begin{eqnarray}
L = - m \bigg( (1 - \frac{x}{x^2 + a^2 \cos^2 \theta}) (t_{, \tau})^2 + 
2 a \frac{x}{x^2 + a^2 \cos^2 \theta} \sin^2 \theta \, (\phi_{, \tau}) (t_{, \tau})
\nonumber\\
- \frac{x^2 + a^2 \cos^2 \theta}{x^2 - x + a^2} (x_{,\tau})^2 -
(x^2 + a^2 \cos^2 \theta) (\theta_{,\tau})^2 - 
\frac{( x^2 + a^2 )^2 - (x^2 - x + a^2) a^2 \sin^2 \theta}{x^2 +
a^2 \cos^2 \theta} \sin^2 \theta (\phi_{, \tau})^2 \bigg)^{1/2} .
\end{eqnarray}

Two of the four Lagrange equations  -- the ones that correspond to the ignorable coordinates $t$ and $\phi$ -- give conservation of full energy $E \equiv m \epsilon$ and momentum $p_{\phi} \equiv m j$:
\begin{eqnarray}
  \label{i10}
E (\equiv m \epsilon) = -\frac{\partial L}{\partial t_{,\tau}} \equiv
(t_{, \tau})^{-1} \bigg( q_{, \tau}^j \frac{\partial L}{\partial q_{, \tau}^j} - L \bigg) = 
- \frac{m^2}{L} \bigg( (1 - \frac{x}{\zeta^2} ) t_{,\tau} + a \frac{x}{\zeta^2} \sin^2 \theta \, \phi_{, \tau } \bigg) 
\end{eqnarray}
and
\begin{eqnarray}
p_{\phi} (\equiv m j) = \frac{\partial L}{\partial \phi_{,\tau}} = \frac{m^2}{L} \bigg(
 a x (t_{, \tau}) - \Lambda \phi_{,\tau} \bigg) \frac{\sin^2 \theta}{\zeta^2} . 
  \label{m}
\end{eqnarray}
Here, the exact form of Lagrangian $L$ is defined above.

Starting from this point and proceeding further on, normalization of 4-velocity permits using the fact that the Lagrangian $L = - m$ (constant) on the trajectory.

Solutions of these equations with respect to $\phi_{, \tau}$ and $t_{, \tau}$ take form
\begin{eqnarray}
\label{tphi}
\phi_{, \tau} = -\frac{4 L}{m \Delta} \bigg(
(x^2 + a^2 \cos^2 \theta )(a^2 j \cot^2 \theta + 
x (a \epsilon + j (-1 + x) \csc^2 \theta ))
\bigg), \nonumber\\
t_{, \tau} = -\frac{4 L}{m \Delta} \bigg(
(x^2 + a^2 \cos^2 \theta )(-a j x + a^4 \epsilon + 
2 a^2 x^2 \epsilon + x^4 \epsilon - a^2 (a^2 + (-1 + x) x) \epsilon \sin^2 \theta)
\bigg),
\end{eqnarray}
where $\Delta = (a^2 + (-1 + x) x)(a^2 + 2 x^2 + a^2 \cos^2 2 \theta)$, and
$L = - m$ (when 4-velocity is normalized).
For large $x \gg 1$, these derivatives become $\phi_{, \tau} \simeq j \csc^2 \theta / x^2$ and $t_{, \tau} \simeq \epsilon + \epsilon / x$.

The remaining two nonlinear Lagrange equations for $x (\tau)$ and $\theta (\tau)$ are
\begin{eqnarray}
\label{lagxtheta}
\frac{d}{d \tau} \frac{\partial L}{\partial x_{,\tau}} - \frac{\partial L}{\partial x }=0, \quad
\frac{d}{d \tau} \frac{\partial L}{\partial \theta_{,\tau}} - \frac{\partial L}{\partial \theta}=0,
\end{eqnarray}
into which Eqs.~(\ref{tphi}) should be substituted.
Solutions of the system of these equations can come only from numerical simulations  
(this article omits such exercise). 
 The end product is certain to look like 
 convoluted patterns of 
trajectories -- esthetically pleasing, great for staring, but unfortunately revealing sparse additional 
insight.

\subsection{Trajectory in Equatorial Surface}
\label{sse45}

Value $\theta (\tau) = \pi / 2$ in Eq.~(\ref{lagxtheta}) assures condition ${\partial L}/{\partial \theta_{,\tau}} = Const$. Thus, a body that starts its journey with initial conditions $\theta_{, \tau} (0) = 0$ and $\theta (0) = \pi / 2$,
always remains at the surface $\theta (\tau) = \pi / 2$. Therefore, $\theta_{, \tau} (\tau) = 0$ for all $\tau$.

To determine the trajectory, either the remaining Lagrange equations Eq.~(\ref{lagxtheta}) need to be solved, or the invariant needs to be used:
\begin{eqnarray}
\label{invL}
(1 - \frac{x}{x^2 + a^2 \cos^2 \theta}) (t_{, \tau})^2 + 
2 a \frac{x}{x^2 + a^2 \cos^2 \theta} \sin^2 \theta \, (\phi_{, \tau}) (t_{, \tau})
\nonumber\\
- \frac{x^2 + a^2 \cos^2 \theta}{x^2 - x + a^2} (x_{,\tau})^2 - (x^2 + a^2 \cos^2 \theta) (\theta_{,\tau})^2 - 
\frac{( x^2 + a^2 )^2 - (x^2 - x + a^2) a^2 \sin^2 \theta}{x^2 + a^2 \cos^2 \theta} \sin^2 \theta (\phi_{, \tau})^2 =1
\end{eqnarray}
i.e.
for the motion at the surface $\theta = \pi / 2$,
\begin{eqnarray}
\label{invLplan}
(1 - \frac{1}{x}) (t_{, \tau})^2 +
\frac{2 a }{x} \, (\phi_{, \tau}) (t_{, \tau})
- \frac{x^2}{x^2 - x + a^2} (x_{,\tau})^2 - 
\frac{( x^2 + a^2 )^2 - (x^2 - x + a^2) a^2}{x^2} (\phi_{, \tau})^2 = 1
\end{eqnarray}
with $\theta_{, \tau}  = 0$ and $\theta = \pi / 2$,
together with Eqs.~(\ref{tphi}), which take form
\begin{eqnarray}
\label{phitplan}
\phi_{, \tau} =  \frac{j( x - 1 )+ a \epsilon}{x (a^2 + (-1 + x) x)}, \quad 
t_{, \tau} = \frac{- a j  + x^3 \epsilon +  a^2 (-1 + x) \epsilon}{x (a^2 + (-1 + x) x)}  .
\end{eqnarray}

For the particular case of $\epsilon = 1$ (parabolic trajectory), the Lagrange equations for variables
$x (\tau )$, $\phi (\tau )$, $t (\tau )$, written in the parametric form via $u (\tau ) = x^{-1} (\tau)$,
take form
\begin{eqnarray}
  \label{vs1}
x_{, \tau} = - \sqrt{u - j^2 u^2 + (a - j)^2 u^3},  \quad 
\phi_{, \tau} = \frac{u^2 (j + a u - j u)}{1 - u + a^2 u^2}, \quad 
t_{, \tau} = \frac{1 - a j u^3 + a^2 u^2 (1 + u)}{1 - u + a^2 u^2}.
\end{eqnarray}
The covariant components are obviously calculated following the rule $v_{\alpha} = g_{\alpha \beta} v^{\beta}$.
For this reason, the module of space-velocity must be calculated using metric tensor, according to the rule
$V^2 = v_{j} v^{j} = g_{i j} v^{i} v^{j}$ with $i,j = 1, 2, 3$.

Note that for the Lagrangian in the form of Eq.~(\ref{i3d}), or Eq.~(\ref{i6}), the 3-velocity in the physically accessible region of space-time, under no circumstances exceeds the speed of light, $V^2 \leq 1$.

In the case of a non-rotating black hole with the Schwarzschild metric, when $a = 0$ (and strictly $1 - j^2 u + j^2 u^2 > 0$, i.e. $j < 2$), the body unavoidably "falls" onto the black hole.

The equations describing the motion, become
\begin{eqnarray}
\label{vs1a}
x_{, \tau} = - \sqrt{u (1 + j^2 (-1 + u) u)}, \quad 
\phi_{, \tau} = j u^2,  \quad 
t_{, \tau} =  \frac{1}{1 - u} .
\end{eqnarray}

When $a \neq 0$ and the test body approaches the surface of ergosphere, $u \rightarrow 1$, we obtain from the point of view of the observer in the "proper" frame
\begin{eqnarray}
  \label{vs2}
x_{, \tau} = - \sqrt{1 - j^2  + (a - j)^2 }, \quad 
\phi_{, \tau} = \frac{1}{a}, \quad 
t_{, \tau} = \frac{1 - a j  + 2 a^2 }{a^2 }.
\end{eqnarray}
From the point of view of the observer at the infinity, at the surface of ergosphere,
\begin{eqnarray}
  \label{vs3}
\dot{x} = - \frac{a^2 \sqrt{1 - j^2  + (a - j)^2 }}{1 - a j  + 2 a^2 }, \quad 
\dot{\phi} = \frac{a}{1 - a j  + 2 a^2 }.
\end{eqnarray}
Here, $\dot{x} = x_{, \tau} / t_{, \tau}$ and $\dot{\phi} = \phi_{, \tau} / t_{, \tau}$ are contra-variant radial and angular components of velocity observed from the infinity.

\section{Parabolic Trajectory}
\label{se5}

\subsection{Classification of Regimes}
\label{sse51}

Let us consider, in more detail, motion of a test body along \emph{parabolic} trajectory at the equatorial surface of a rotating black hole with rotation parameter bounded by the condition $-0.5 < a < 0.5$. When there is no motion at the infinity, parameter $\epsilon = 1$. To find trajectory $x =x (\phi)$, one should make transformation $x = u^{-1}$  and account for $x_{, \tau} = - \phi_{, \tau} u^{-2} u_{, \phi}$.

After calculations, we obtain the trajectory equation for $x^{-1} = u (\phi)$:
\begin{eqnarray}
\label{eqplan0}
(\frac{d u}{d \phi})^2 = \frac{(1 - u + a^2 u^2)^2}{(j + a u - j u)^2} u \bigg( 1 - j^2 u +(a - j)^2 u^2 \bigg) .
\end{eqnarray}

For the newtonian approximation, we must put parameter $a \rightarrow 0$ and neglect the small term $\sim u^2$ in the brackets of the right part of Eq.~(\ref{eqplan0}). Then the solution is
$u (\phi) = (2 j^2)^{-1} (1 - \cos \phi )$,
i.e., the body moves along the parabolic trajectory around the gravitating center. For large $j \gg 1$
(for practical estimations, $j > 4 \div 5$), quantity $u \leq u_{max} = (4 j^2)^{-1} \ll 1$.

For small $\phi^2 \ll 1$, and consequently, for small $u$ (large $x = u^{-1}$), estimation
$\phi (u) \simeq \pm 2 j \sqrt{u}$ is valid.

In the case of a rotating black hole, when $a \neq 0$, the event horizon and the surface of ergosphere are separate, and 
two regimes of motion are possible for the traveling body.

In the first regime, the body starts its motion at the infinity, where parameter $u=0$, with zero initial velocity, i.e.,
$\epsilon = 1$. The body then arcs around the rotating black hole, approaching within the minimal distance $x_{min}^{-1} = u_1$, and then leaves to infinity.
The movement of the body occurs, first, in the regime from $0$ to $u_1$,
and then from  $u_1$ to $0$. This scenario is possible if condition $j^4 > 4 (j- a)^2$ is satisfied in expressions
\begin{eqnarray}
\label{korni12}
u_{1,2}=\frac{j^2 \mp \sqrt{-4 a^2 + 8 a j - 4 j^2 + j^4} }{2 (a^2 - 2 a j + j^2)}, \; u_3 = \frac{j}{j - a} .
\end{eqnarray}
The trajectory equation can then be written as
\begin{eqnarray}
\label{eqplan}
\frac{d \phi}{d u} = \pm \frac{u_3 - u}{(1 - u + a^2 u^2) \sqrt{u (u_1 - u) (u_2 - u)}} .
\end{eqnarray}
If the body moves from infinity towards the black hole, i.e., from $u = 0$ to $u = u_1$ for the flyover trajectory,
the motion is described by the branch with the plus-sign.

In the opposite direction -- from $u = u_1$ to $u = 0$ -- the sign is minus. For flyover trajectories, $u_1 < u_3$ always.

Definition~(\ref{korni12}) can be rewritten as
\begin{eqnarray}
\label{korni2}
u_{1,2}=\frac{u_3^2}{2} ( 1 \mp \sqrt{ \delta} ), \; u_3 = \frac{j}{j - a}, \; \delta = 1 - \frac{4}{j^2 u_3^2}.
\end{eqnarray}
Roots are real if $\delta > 0$.

\begin{figure}[h!]
 \centering
 \includegraphics[width=8cm]{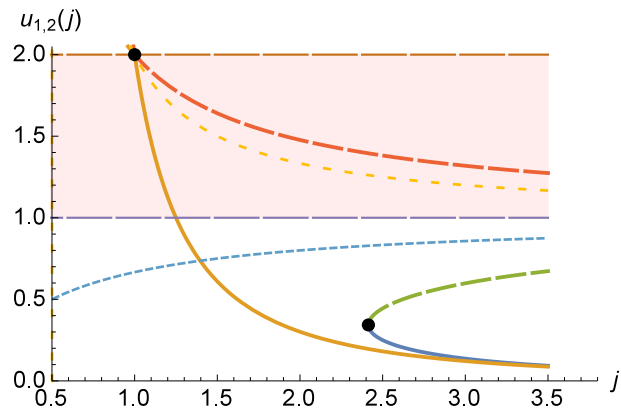}
 \caption{
 Roots $u_{1,2}$ as functions of initial specific momentum (angular momentum per mass unit) $j$.
 Dotted lines depict $u_3^{\pm}$. Pink zone is the ergosphere. 
 }
 \label{korniFig}
\end{figure}

Quantities $u_{1,2}$ as functions of parameter $j$ are shown in Fig.~\ref{korniFig}. In this Figure, solid lines show the branches of roots $u_1^-$ (upper) and 
$u_1^+$ (lower) that correspond to the limit cases of the rotation parameter $a = \mp 0.5$. Dashed lines show $u_2^-$  and $u_2^+$. At the points marked by large dots, the roots merge, $u_{1,2}^{\mp} = {(u_3^{\mp})^2} / {2}$. The physical, external region is defined by the condition $0 \leq u \leq 1$
(from the infinity $u = 0$ to the static limit $u = 1$). Line $u_S = 1$ is the static limit.
Line $u_H = ( \sqrt{1 - 4 a^2} / 2 )^{-1} = 2$ is the event horizon. The region between $u_S = 1$ and $u_H = 2$ represents the ergosphere.

When $a = - 0.5$ (the black hole rotates clockwise), the body with the angular momentum per mass unit
 $j < j_m = 1 + \sqrt{2} \simeq 2.414$ "pierces" 
 the 
 static limit and reaches the event horizon, i.e., the body will be captured by the black hole. In addition, at the level $u_3^- = j / (j + 0.5)$, the sign of the derivative
 $\phi_{, u}$ changes (level  $u_3^-$ is  shown in the Figure by the lower dotted curve).
 In other words, the body approaching the black hole will be unavoidably involved into co-rotation with the black hole.

When $a = + 0.5$ (rotation of the black hole is counter-clockwise), the body with specific angular momentum $1 < j < j_p =1.25$
"dives" under the static limit surface into the ergosphere, reaches the closest distance with the black hole $u_1^{-1}$,
 and "dives out" to proceed on its way to infinity.

For $a = + 0.5$, the body will be captured by the black hole only for small $j \leq 1$. No change of sign of derivative $\phi_{, u}$ happens in this scenario.

At the relatively large values $j \gg 1$ (typically, $j > 3$), the roots may be calculated using formulae $u_1 \simeq j^{-2} + j^{-4}$, $u_2 \simeq 1 + 2 a / j - (1 - 3 a^2)/j^2$, and $u_3 \simeq 1 + a / j - a^2 / j^2$.

\begin{figure}[h!]
 \centering
 \includegraphics[width=8cm]{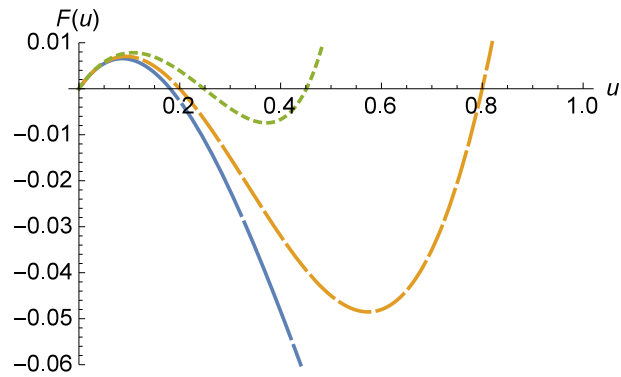}
 \caption{
The right part of Eq.~(\ref{eqplan}) with $j=2.5$ for values $a = 0$ (middle line, medium dashes), $a = 0.49$ (lower line, long dashes) and $a = - 0.49$ (upper line, short dashes). Motion of test body is possible in the domains where the root branches have positive values.
}
 \label{Proizv}
 \end{figure}

The characteristic behavior of the test body for a relatively large value of parameter $j$, when $\delta > 0$,
may be qualitatively understood based on Fig.~\ref{Proizv}. For illustration, the calculations use $j= 2.5$.
In this case, real roots $u_{1,2}$ exist. Then, having started at the infinity with $u = 0$ with zero initial velocity, the test body reaches the closest distance from the black hole (defined by the smallest of the real roots of $u_1$)
and returns to infinity, $u = 0$, at the angle  $\phi \neq 2 \pi$.

 \begin{figure}[h!]
 \centering
 \includegraphics[width=8cm]{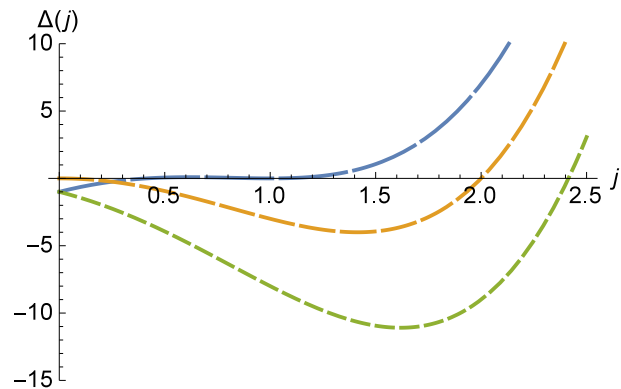}
 \caption{Determinant
 $\Delta (j)= - 4 a^2 + 8 a j - 4 j^2 + j^4 $
 as function of $j$ for values $a = 0.5$ (blue upper line), $0$ (yellow middle line), and $a = -0.5$ (green lower line).
 }
 \label{delta}
 \end{figure}

\begin{figure}[h!]
\centering
\includegraphics[width=8cm]{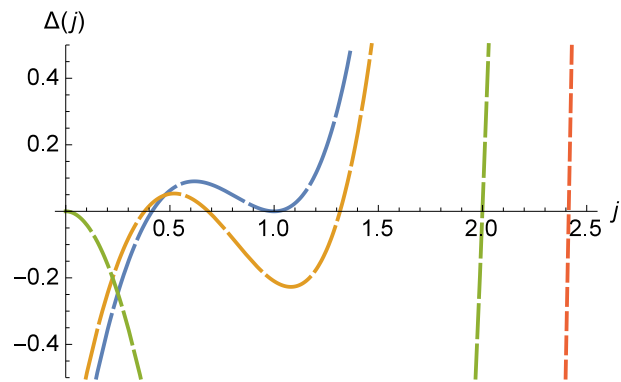}
\caption{Zoom into Fig.~\ref{delta} showing the same determinant $\Delta (j)= - 4 a^2 + 8 a j - 4 j^2 + j^4 $ for values $a = 0.5$ (blue upper line), $0.45$ (yellow line), $a = 0$ (green line), and $a = -0.5$ (red line). It is apparent that with a certain combination of parameters  -- $j$-rotation of the test body around the black hole and proper $a$-rotation of the black hole -- there arises a "window of transparency" where $\Delta (j)> 0$. However, for such window $u_1 > 1$ and the noted flyover regime does not become realized in the region $0 \leq u \leq 1$, i.e., outside the ergosphere.
}
\label{delta2}
\end{figure}

Another scenario takes place when $\delta < 0$ and $0 < j < j_{m,p}$, which is equivalent to $\Delta = -4 a^2 + 8 a j - 4 j^2 + j^4  <0$, i.e., when real roots $u_{1,2}$ do not exist. Then, the approaching body  unavoidably becomes captured by the black hole and falls onto the  static limit surface $u = 1$, i.e., onto the boundary of the ergosphere.

\begin{figure}[h!]
    \centering
    \includegraphics[width=7cm]{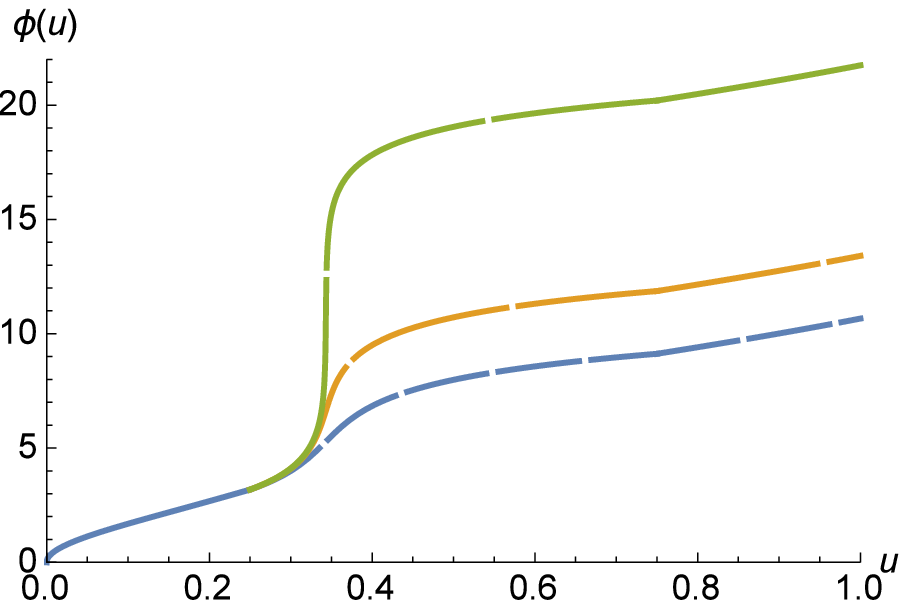}
    \caption{
    Coordinate $\phi [u]$ as function of $u$ as the body falls onto the fast-rotating black hole for parameters $a = - 0.5$ and $j = 1 + \sqrt{2} - \beta$ where $\beta = 10^{-2}$ (lower), $10^{-3}$ (middle) and $10^{-6}$ (upper).
    }
    \label{phiNcrit}
\end{figure}

For $j = 1+\sqrt{2} - \beta$ when $\beta \rightarrow + 0$,
near $u \simeq 0.343$, change of angle $\phi$ occurs most rapidly (Fig.~\ref{phiNcrit}).
This is where the majority of the spins occurs.

\begin{figure}[h!]
    \centering
    \includegraphics[width=6.5cm]{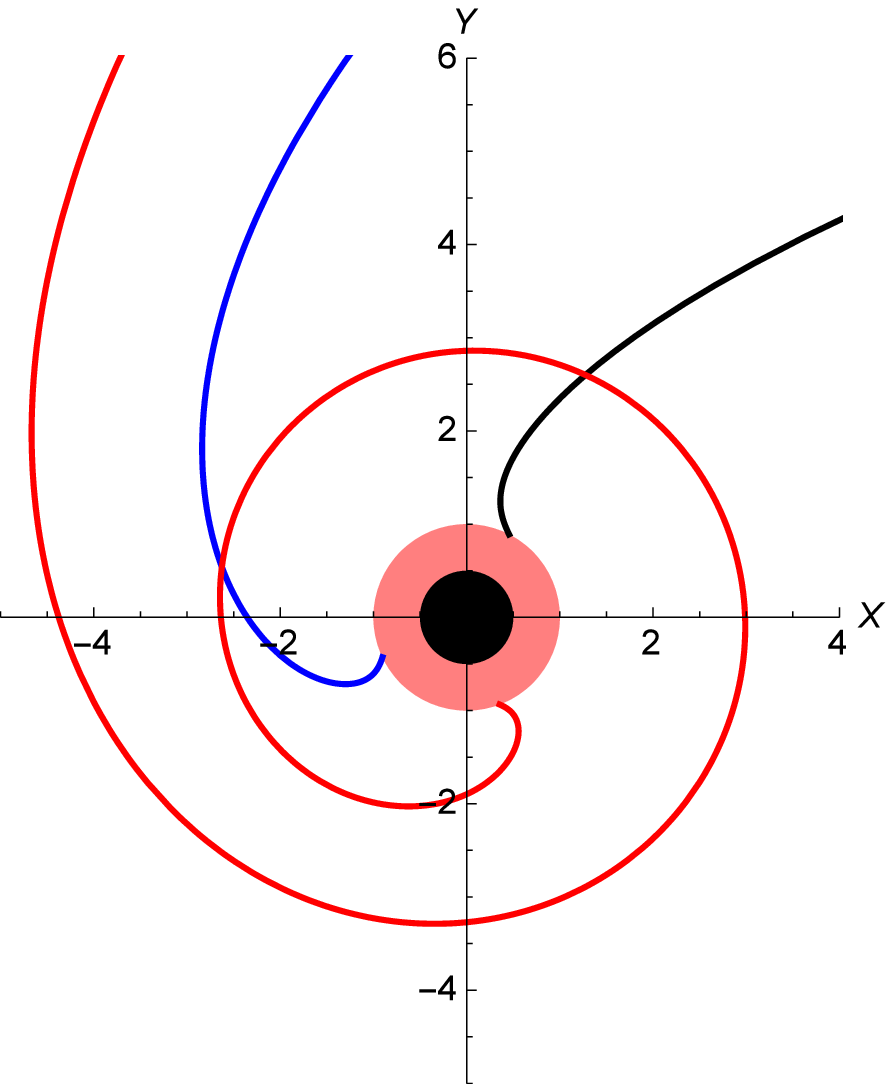}
    \caption{
  Numerical simulation of the scenarios where the traveling body approaches the black hole (rotating in the opposite direction) and becomes entrapped by the "whirlpool" of space-time. 
  }
    \label{zakhvat}
\end{figure}

Falling onto the static limit occurs either (if $a >0$, when $u_3 > 1$)
along the inward spiral without any peculiarities, or (if $a < 0$) with the change of the direction of rotation of the body around the black hole at level $u_3 < 1$
(when $d \phi / d u < 0$) .
In particular, after traversing level $u = u_3$ and further moving towards level $u = 1$,
the body starts moving in the same direction as the black hole's rotation.
The numerical simulation of this scenario is shown in Fig.~\ref{zakhvat}.  During the capture process, the trajectory sharply curves and the body starts co-rotating with the black hole.
  In the figure, the black hole (at the center) rotates with  $a = - 0.5$.  Its radius (maximum permitted for such rotation) equals to $0.5$ (for the plane $\theta = \pi / 2$). Radius for ergosphere (marked in pink) equals to $1$. Three trajectories are shown for $j = 1$ (black), $j = 2$ (blue), and $j \simeq 1 + \sqrt{2} - 10^{-3}$ (red).

\begin{figure}[h!]
    \centering
    \includegraphics[width=7.5cm]{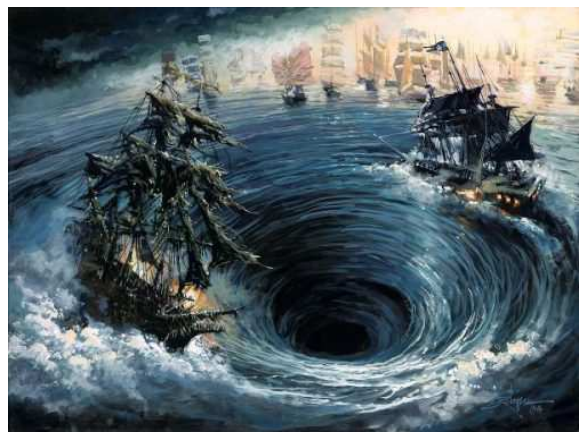}
    \caption{
    Ocean whirlpools in legends.  Illustration by unknown artist.
    }
    \label{vodovorot}
\end{figure}

The impact of the curved space-time (the Kerr metric) in the vicinity of the black hole remarkably resembles a gigantic whirlpool (Fig.~\ref{vodovorot}).\footnote{
A whirlpool is a rotating current of water which creates a characteristic vortex.
In nature, there exists a number of consistent and frequently occurring whirlpools that have become legendary.
Many sea myths 
have featured whirlpools, typically in situations involving great peril to ships. 
An especially powerful whirlpool is known as a maelstr\"{o}m -- one of the more notable maelstr\"{o}ms is the Moskstraumen, an immense network of eddies and whirlpools off the coast of Norway.}

Just like an ocean whirlpool swirls, traps and swallows a boat  -- a situation dramatically described in Edgar Allan Poe's short story \emph{A Descent into the Maelstr\"{o}m}) \citep{p41} -- so does the black hole.  When the "aiming" parameter $j$ is less than the certain critical value $j_*$ (for $a = + 0.5$, quantity $j_* \equiv j_+ = 1.25$, and for $a = - 0.5$, quantity $j_* \equiv j_- = 1 + \sqrt{2}$), the approaching object cannot free itself once entrapped.

\subsection{
At what angle does the body come closest to the black hole?}
\label{sse52}

The answer to the question 'At what angle does the body come closest to the black hole?' follows from integrating Eq.~(\ref{eqplan}) for a flyover trajectory, when
$\delta > 0$ (and keeping in mind that $\sqrt{t^2} = |t|$):
\begin{eqnarray}
\label{eqplan2}
\phi_m = \int_0^{u_1} d \xi \frac{u_3 - \xi}{\sqrt{\xi (u_1 - \xi)(u_2 - \xi)}} K ( \xi ) ,
\end{eqnarray}
where
\begin{eqnarray}
\label{eqK}
K ( \xi ) = (1 - \xi + a^2 \xi^2)^{-1/2}  .  
\end{eqnarray}

\begin{figure}[h!]
    \centering
    \includegraphics[width=8cm]{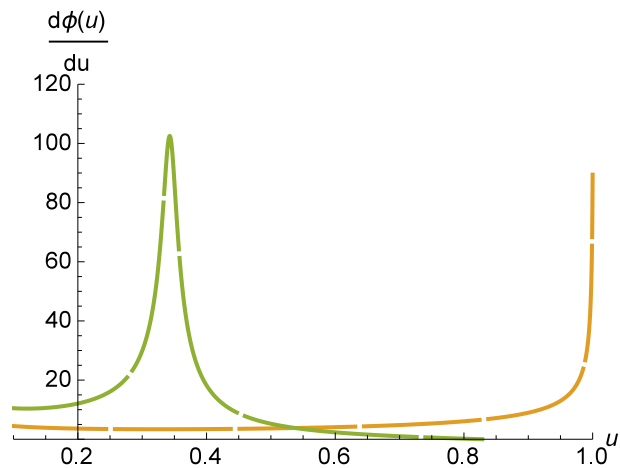}
    \caption{
    Derivative $\phi_u$ as function of $u$ for limit parameters $a = \pm 0.5$, i.e., for
    $j_{+} = 1 + \sqrt{2}$ and $j_{-} = 1.25$. For $a = - 0.5$ (left curve), domain $u_1 = 0.343 \leq u \leq 1$ is not accessible by the test body.
    }
    \label{proiz}
\end{figure}

Due to the presence of $K(\xi)$, the integral cannot be calculated analytically.  However, in the vicinity of level $u_*$
where the flyover regime switches to the capture regime, the main contribution into the integral in Eq.~(\ref{eqplan2})
comes from the domain near the boundary $u_*$ : $u_{*,-} \rightarrow 0.34$ and $u_{*,+} \rightarrow 1$. This is where
(as seen in Fig.~\ref{proiz})
the fastest change in derivative $d \phi / d u$ happens and the sharp twisting of the trajectory occurs. In this domain, an approximation for $K ( \xi )$ can suffice.

Let's expand $K ( \xi )$ into a series  in the vicinity of $u_1$. Even the first few terms of the expansion offer an acceptable approximation for the entire region of changing $u$

\begin{eqnarray}
\label{eqK2}
K ( \xi ) = \sum_{k=0}^{\infty} \frac{b_k}{Z^{k+1}} (\xi - u_1)^k =
\nonumber\\
\frac{1}{Z}  +
\frac{1 - 2 a^2 u_1}{Z^2} (\xi - u_1) + 
\frac{1 - a^2 - 3 a^2 u_1 + 3 a^4 u_1^2}{Z^3} (\xi - u_1)^2 + 
\frac{b_3}{Z^4} (\xi - u_1)^3 +
\frac{b_4}{Z^5} (\xi - u_1)^4 + ...   .
\end{eqnarray}
Here, $Z = 1 - u_1 + a^2 u_1^2 $, $b_3 = 1 - 2 a^2 - 4 a^2 u_1 + 4 a^4 u_1 + 6 a^4 u_1^2 - 4 a^6 u_1^3$, $b_4 = 1 - 3 a^2 + a^4 - 5 a^2 u_1 + 10 a^4 u_1 + 10 a^4 u_1^2 - 10 a^6 u_1^2 - 10 a^6 u_1^3 + 5 a^8 u_1^4$, etc.

Eq.~(\ref{eqplan2}) can then be written in the form

\begin{eqnarray}
\label{eqplan3}
\phi_m = \sum_{k=0}^{\infty} b_k \bigg( 2 \frac{\partial}{\partial u_2} \int_0^{u_1} d \xi \frac{\sqrt{u_2 - \xi}}{\sqrt{\xi (u_1 - \xi)}} (\xi - u_1)^k \bigg).
\end{eqnarray}
The auxiliary integral
\begin{eqnarray}
\label{eqplan3b}
 I_k (u_1, a) =  \int_0^{u_1} d \xi \frac{\sqrt{u_2 - \xi}}{\sqrt{\xi (u_1 - \xi)}} (\xi - u_1)^k
\end{eqnarray}
and its doubled derivative $ 2 I_k ' (u_1, a) \equiv f_k$ can be easily calculated here, yielding a simple
\footnote{
"Simplicity", of course, is a \emph{relative} concept, on par with "large", "small" and "all we need", as humored by Mark Twain, a titan of the American literature:
"... My friend, take an old man's advice, and don't encumber yourself with a large family--mind, I tell you, don't do it. In a small family, and in a small family only, you will find that comfort and that peace of mind which are the best at last of the blessings this world is able to afford us, and for the lack of which no accumulation of wealth, and no acquisition of fame, power, and greatness can ever compensate us. Take my word for it, ten or eleven wives is all you need--never go over it."
}
analytical expression for the leading terms of the expansion 

\begin{eqnarray}
\label{fk}
f_k =
\frac{\sqrt{\pi} \Gamma(\frac{1}{2} + k)}{2 (1 + k) \Gamma(1 + k) \sqrt{- u_1 + u_2}} (- u_1)^k 
\bigg[
-\bigg( 2 (1 + k) \, _2F_1 (-\frac{1}{2}, \frac{1}{2} + k, 1 + k, \frac{u_1}{u_1 - u_2})
\nonumber\\
- (1 + 2 k) \, _2F_1 (-\frac{1}{2}, \frac{3}{2} + k, 2 + k, \frac{u_1}{u_1 - u_2})
\bigg) u_1 
+ 2 (1 + k) \, _2F_1 (-\frac{1}{2}, \frac{1}{2} + k, 1 + k, \frac{u_1}{u_1 - u_2}) u_3
\nonumber\\
- \frac{(1 + 2 k) u_1}{2(2 + k)(u_1 - u_2)} 
\bigg(
\bigg(
2 (2 + k) \, _2F_1 (\frac{1}{2}, \frac{3}{2} + k, 2 + k, \frac{u_1}{u_1 - u_2}) 
\nonumber\\
- (3 + 2 k) \, _2F_1 (\frac{1}{2}, \frac{5}{2} + k, 3 + k, \frac{u_1}{u_1 - u_2})
\bigg) u_1 
- 2 (2 + k) \, _2F_1 (\frac{1}{2}, \frac{3}{2} + k, 2 + k, \frac{u_1}{u_1 - u_2}) u_3 \bigg) .
\bigg]
\end{eqnarray}
Here, $\, _2F_1 ( \alpha , \beta , \gamma , z )$ is the hypergeometrical function of the second kind, $\Gamma (z)$ is the gamma--function.

The main contribution comes from the leading term

\begin{eqnarray}
\label{eqI01} 
f_0 = \frac{\pi}{2 \sqrt{-u_1 + u_2}}
\bigg[
\bigg(
-\frac{4}{\pi} E ( \frac{u_1}{u_1 - u_2} ) + 
\,_2F_1 ( -\frac{1}{2}, \frac{3}{2}, 2, \frac{u_1}{u_1 - u_2} ) \bigg) u_1
+
\frac{4}{\pi} E ( \frac{u_1}{u_1 - u_2} ) u_3 -
\nonumber\\
   \frac{u_1}{4 (u_1 - u_2)}
   \bigg(
   (
   4 \,_2F_1 ( \frac{1}{2}, \frac{3}{2}, 2, \frac{u_1}{u_1 - u_2} ) - 
   3 \,_2F_1 ( \frac{1}{2}, \frac{5}{2}, 3, \frac{u_1}{u_1 - u_2} )
   ) u_1 - 
4 \,_2F_1 ( \frac{1}{2}, \frac{3}{2}, 2, \frac{u_1}{u_1 - u_2} ) u_3 .
\bigg)
\bigg]
\nonumber\\
\end{eqnarray}

It is worth reminding that in the equatorial plane, the boundary of the event horizon is determined by the value
$u_H \equiv (x_H)^{-1} = 2 (1 + \sqrt{1 - 4 a^2})^{-1}$, and the surface of the static limit -- by $u_E = 1$. Therefore, for a physically realizable trajectory, it always holds that $u \leq u_1 < 1$.

\begin{eqnarray}
\label{eqUgol}
\phi_m \simeq \frac{\pi}{2 (1 - u_1 + a^2 u_1^2 )\sqrt{-u_1 + u_2}} 
\bigg[
\bigg(
-\frac{4}{\pi} E ( \frac{u_1}{u_1 - u_2} ) + 
\,_2F_1 ( -\frac{1}{2}, \frac{3}{2}, 2, \frac{u_1}{u_1 - u_2} ) \bigg) u_1 +
\nonumber\\
\frac{4}{\pi} E ( \frac{u_1}{u_1 - u_2} ) u_3 - 
\frac{u_1}{4 (u_1 - u_2)} \bigg((
   4 \,_2F_1 ( \frac{1}{2}, \frac{3}{2}, 2, \frac{u_1}{u_1 - u_2} ) - 
3 \,_2F_1 ( \frac{1}{2}, \frac{5}{2}, 3, \frac{u_1}{u_1 - u_2} )) u_1 -
\nonumber\\
4 \,_2F_1 ( \frac{1}{2}, \frac{3}{2}, 2, \frac{u_1}{u_1 - u_2} ) u_3
\bigg)
\bigg]
\end{eqnarray}

Recall that parameters $u_{1,2}$ are
\begin{eqnarray}
\label{ubeta}
u_{1,2}=\frac{u_3^2}{2} (1 \mp \sqrt{\delta} ), \; u_3 = \frac{j}{j - a}
, \, \delta = 1 - \frac{4}{j^2 u_3^2} . 
\end{eqnarray}

Solution for the Schwarzschild metric results from Eq.~(\ref{eqUgol}) by setting $a=0$, i.e., $u_3 = 1$.

The newtonian limit results from making approximation with $a = 0, j \gg 1$,
i.e., from formally setting in Eq.~(\ref{eqUgol}) $u_1 \rightarrow 0$, $u_2 \rightarrow 1$ and $u_3 \rightarrow 1$.
This produces the well-expected result:  if a body starts at infinity with initial $\phi = 0$ and zero-velocity, then at the moment of maximum approach with the gravitating center, the position of the body is characterized by $\phi_m = \pi$.

\subsection{How many rotations does the body complete before departing towards infinity?}
\label{sse53}

To determine how many rotations the body completes before departing toward infinity, we need to calculate the integral along both branches of the trajectory -- the approach branch with $u_{, \tau} > 0$ and the departure branch with $u_{, \tau} < 0$:
\begin{eqnarray}
\label{nombre1}
N = \frac{1}{2 \pi} \int_0^{u_1} d \xi \frac{u_3 - \xi}{\sqrt{\xi (u_1 - \xi)(u_2 - \xi)}} - 
\frac{1}{2 \pi} \int_{u_1}^0 d \xi \frac{u_3 - \xi}{\sqrt{\xi (u_1 - \xi)(u_2 - \xi)}} .
\end{eqnarray}
By substituting $\xi \rightarrow u_1 - \xi$, the second integral becomes
\begin{eqnarray}
\label{nombre2}
N = 
\frac{1}{2 \pi} \int_0^{u_1} \frac{d \xi}{\sqrt{\xi (u_1 - \xi)}} \bigg(
\frac{u_3 - \xi}{\sqrt{u_2 - \xi}} +
\frac{u_3 - u_1 + \xi}{\sqrt{u_2 - u_1 + \xi)}} \bigg).
\end{eqnarray}
The main contribution to the integral comes from the domain near $u_1$.

\begin{figure}[h!]
\centering
\includegraphics[width=8cm]{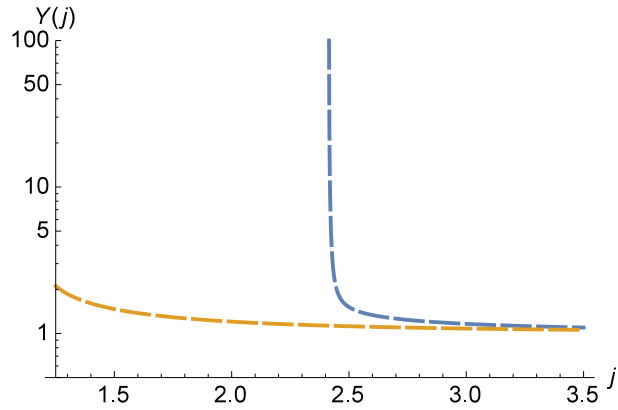}
\caption{
Number of turns $N$ as function of specific angular momentum $j$ when test body travels in the vicinity of fast-spinning black hole, shown for two rotation regimes, $a = - 0.5$ (retrograde, right curve) and $a = + 0.5$ (prograde, left curve).
}
\label{Numbj}
\end{figure}

The number of rotations, as a function of $j$, is shown in Fig.~\ref{Numbj}.

In a particular example, at the maximum value allowed by the theory of black holes, $a = + 0.5$ (when both the black hole and the test  body rotate in the same direction, clockwise in our analysis) and when $j = 1.249$, the number of rotations is $N \simeq 2.12$. In other words, upon arriving from the point with coordinates $x= \infty, \, \phi = 0$, and after completing two rotations around the black hole, the body departs towards the infinitely-remote point with coordinates $x = \infty , \, \phi = 0.12$ radian.
When the black hole rotates counter-clockwise, with $a = - 0.5$ and $j \simeq 2.4150 > 1 + \sqrt{2} \simeq 2.41421$
(note the difference from the critical value is only in the forth decimal),
the body completes more than a hundred rotations, $N \simeq 102.76$.
In other words, upon arriving from the point with coordinates $x = \infty, \, \phi = 0$ and completing one hundred and two rotations around the black hole, the body departs towards the point with coordinates $x = \infty , \, \phi = 0.76$ radians.

If $j \rightarrow 1 + \sqrt{2}$, the rise in the number of rotations happens rapidly, $N \rightarrow \infty$.  The trajectory of the body "spirals" infinitely onto the minimally close circular orbit defined by $u_1 \simeq 0.343$, to start "unwinding later, after infinite number of turns".

\begin{figure}[h!]
\centering
\includegraphics[width=8cm]{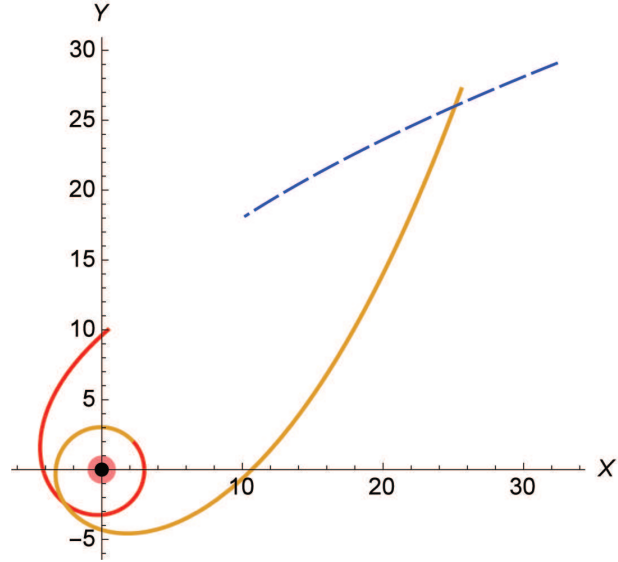}
\caption{
Overfly trajectory near retrograde-rotating black hole for initial conditions 
$a = - 0.5$ and $j = 1+ \sqrt{2} + 10^{-3}$.
These parameters were selected to reveal clearly on the graph the impact of the black hole rotation on the trajectory of the traveling test body. 
Dashed line sketches the initial parabolic part of the test body trajectory, as the body approaches the retrograde--spinning black hole. 
}
\label{ProlTraj}
\end{figure}

Fig.~\ref{ProlTraj} presents a typical trajectory of the body approaching a clockwise-rotating black body with the maximum allowed value $a = - 0.5$ and the value of specific angular momentum $j = 1+ \sqrt{2} + 10^{-3}$
only slightly exceeding the "capture" limit $j_* = 1+ \sqrt{2}$. The body, arriving along the initially parabolic trajectory from the initial location characterized by $x \rightarrow \infty$ and $\phi_{in} \rightarrow 0$ (following the dashed line from right to left),
then curves around the black hole, gets involved in the rotation of the space-time in the vicinity of the black hole, and flies away towards infinity with the angle (relative to the axis) $\phi_{fin} \neq 2 \pi$.

\section{Explicit Lagrangian in Post--Newtonian Approximation}
\label{se5}

The Lagrangian written for the test body moving in the space--time $q^{\mu} = (t, x, \theta , \phi)$ as seen by a distant observer, is written as Eq.~(\ref{i3d}). 
Dots over variables represent derivatives \emph{with respect to time $t$ measured by the observer located at the infinity}.

Consideration of Lagrangian~(\ref{i6}) in its general form  is not particularly informative because the avalanche of numerical details buries the insights. 
But certain limit cases appear useful.

To grasp the impact of the black hole rotation on the motion of the body, we restrict consideration to the motion in the equatorial surface of the black hole, $\theta = \pi / 2$.

First, let us remind that from Eqs.~(\ref{i5}), it follows that  $\theta = \pi / 2, \; \dot{\theta} = 0$  satisfy the equations of motion 
-- if the motion is in the equatorial plane initially, it always remains in the plane.
This result follows from the uniqueness theorem for solutions of differential equations. With $\theta = \pi / 2$ and $\dot{\theta} = 0$, the Lagrangian is

\begin{eqnarray}
  \label{i6c2}
L = - m \bigg[ 1 - \frac{1}{x} + \frac{2 a}{x} \dot{\phi}
  - \frac{x^2}{x^2 - x + a^2} \dot{x}^2 
  - \frac{1}{x^2} \bigg( ( x^2 + a^2 )^2 -
  (x^2 - x + a^2) a^2 \bigg) \dot{\phi}^2 \bigg]^{1/2} .
\end{eqnarray}

It gives correct limit expression when the observer is at a large distance from the black hole, i. e. when $x \rightarrow \infty$ and the gravity field produced by the black hole can be neglected. In this case, $L \rightarrow - m \sqrt{1 - v^2}$ where $\dot{x}^2 + x^2 \dot{\phi}^2 = v^2$. This is exactly the Lagrangian of a free (relativistic) test particle in the absence of gravitational field \citep{ll69}.

For a relativistic particle (with $\dot{x}^2 + x^2 \dot{\phi}^2 \equiv v^2 < 1$, $x > 1$, $a^2 <1$),
by keeping the leading terms in expansion~Eq.~(\ref{i6c2}) -- in particular, by keeping only the permanent terms with $x^{-2}$,
for $x^{-1}$ keeping only the terms with no greater than second order with respect to velocity, for the terms containing only the velocity keeping the terms not greater than the fourth order with respect to $v$, and finally, keeping the term linear in $a$ -- we obtain that
\begin{eqnarray}
\label{i6g}
L = L_0 + L_1 \simeq
\nonumber\\ 
 \bigg[ - m + \frac{m v^2}{2}  + \frac{m}{2 x} \bigg] + 
\bigg[
\underbrace{\frac{m}{8} ( \dot{x}^2  + x^2 \dot{\phi}^2)^2} +
\underbrace{\frac{m}{8 x^2}} +
\underbrace{\frac{m}{4 x} ( 3 \dot{x}^2 + x^2 \dot{\phi}^2 )} - 
\underbrace{- \frac{m}{x} a \dot{\phi} ( 1 + \frac{1}{2} x^2 \dot{\phi}^2 )} 
\bigg] .
\end{eqnarray}
The first term in $L_1$ of Eq.~(\ref{i6g})  is the correction that follows from the special theory of relativity -- when the velocity of the particle is not small relative to the speed of light.
The second term follows from the general relativity theory (the case of the Schwarzschild metrics when $a = 0$).
The next term may be interpreted as the result of interaction of the gravitational field created by the central body with the moving object.
Curiously, this term is anisotropic along the radial and transversal components of velocity.
The last term, which depends on the direction of the black hole rotation, follows from the Kerr
metrics near the black hole.
Coordinate $x$ here is interpreted as "distance" from the black hole, even though, as clarified in Introduction, it is not quite so.

The classical Newtonian approximation corresponds to the case when $L_1$ is neglected.

The constant $m$ in Lagrangian~(\ref{i6g}) does not became apparent in the equations of motion.
However, this constant is important in the expression for the energy. The full energy following from
\begin{eqnarray}
  \label{energ}
E = u^{j} \frac{\partial L}{\partial u^{j}} - L \, ,
\end{eqnarray}
where $j=1,2,3$ are space indices, is
\begin{eqnarray}
  \label{i7e}
E = m + \frac{m}{2} ( \dot{x}^2  +
  x^2 \, \dot{\phi}^2 ) - \frac{m}{2 x}  + \, ... \,  .
\end{eqnarray}
These expressions explain the meaning of the parameter $m$. Being measured in energetic units ($m
\rightarrow m c^2$), it gives the rest--mass energy of the particle, $E_0 = m c^2$.
Expression~(\ref{i7e}) contains the rest-mass energy of the test particle, its kinetic energy, and the potential
energy, of the particle in the central gravitational field. 
For Lagrangian,  the form of Eq.~(\ref{i3d}) is well chosen because the square-root form is what yields the correct coefficient 
$1/2$ in the expression for potential energy in the classical Newtonian approximation  Eq.( \ref{i7e}),  because  $r_g = 2 G M_h / c^2$  in $x = r / r_g$  is defined with coefficient $2$.

However, in the post--newtonian approximation, everything evolves as if in the usual classical space-time, the Lagrangian $L_0$ gained an additional term, $L_0  \rightarrow L_0 + L_1$,
leading to appearance of additional field forces, $\partial_{q^j} L_1$.
Let's notice that the second term in $L_1$ is responsible for the appearance of the additional force -- in a certain sense, an analogue of the Coriolis or Lorentz force -- because $L_1 \sim (\mathbf{\nabla} U_{newt} \cdot [\mathbf{J} , \mathbf{v}])$ with $U_{newt} = - m / 2 x$.
Notably, for this effect to arise, it is necessary that both, $\partial_{x x} U_{newt}$, $\mathbf{J}$ and $\mathbf{v}$, are non-zero simultaneously.

The described model can be useful for analysis of problems when extended, elastically-deformable, plasma objects
(with or without a hard "shell") are torn apart,
or when "super-droplets" of super--dense (nuclear) matter travel in the vicinity of astrophysical objects with strong gravitational fields.
If it is known a~priori that an elastic  body 
(not a "dust cloud" composed of 
gravitating particles, for which the Roche limit is derived)
upon its approach to the black hole becomes torn apart into a small number of fragments--droplets,
then to describe the end-result,  there is no need to turn to the hydrodynamical approximation.
Instead, the Lagrangian technique is more natural and better suited.
The method presented in this article is much simpler for numerical simulations and
generates results that are easier to interpret.

Generalization of the proposed model for the case of several "connected" bodies is not complicated.
To study the effect of how, in the post-newtonian approximation, the black hole's rotation impacts the approaching body's trajectory
(in the plane $\theta = \pi / 2$) and how the structure of $N$ bodies with interaction $U_{int}$
(or without one, in the cases of 
"dust cloud" 
or "droplet" without self-gravitation) becomes deformed during the approach,
it is natural to suggest a model Lagrangian in the form

\begin{eqnarray}
  \label{LagFin}
L_N = \sum_{i=1}^N L^{(i)} + \frac{1}{2} \sum_{i=1}^N \sum_{k = 1}^{N \prime} U (r_{ik}) ,
\end{eqnarray}
where
$U (r_{ik})$ is the potential energy of interaction of $i-th$ and $k-th$ bodies, prime signifies absence of diagonal terms in the double sum, and where
\begin{eqnarray}
\label{Lagn}
L^{(i)} = \bigg[ - m_i + \frac{m_i}{2} \bigg( \dot{x_i}^2 + x_i^2 \dot{\phi_i}^2 \bigg)  + \frac{m_i}{2 x_i} \bigg] +
 \nonumber \\
\bigg[
\frac{m_i}{8} ( \dot{x}_i^2  + x_i^2 \dot{\phi}_i^2)^2 +
\frac{m_i}{8 x_i^2} +
\frac{m_i}{4 x_i} ( 3 \dot{x}_i^2 + x_i^2 \dot{\phi}_i^2 ) - 
\frac{m_i}{x_i} a \dot{\phi}_i ( 1 + \frac{1}{2} x_i^2 \dot{\phi}_i^2 ) \,
\bigg] .
\end{eqnarray}
In Eq.~(\ref{Lagn}), it is assumed that $\sum m_i \ll M_h$.

\begin{figure}[h!]
\centering
\includegraphics[width=8cm]{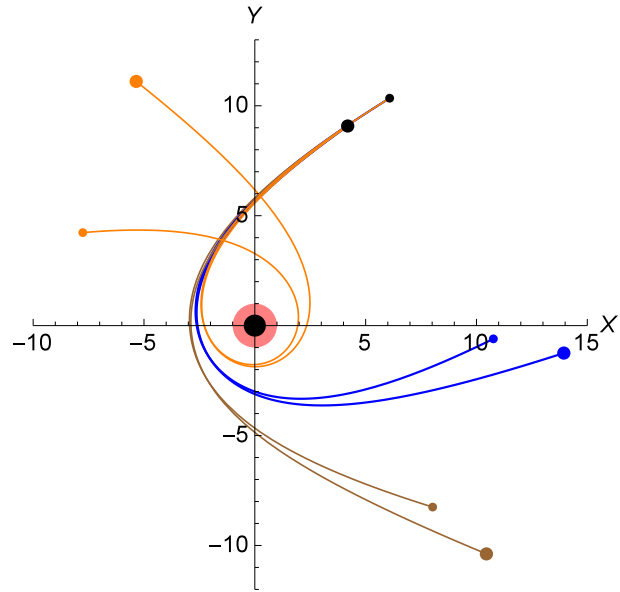}
\caption{
Involvement into co--rotation with space-time "whirlpool", and subsequent destruction into two fragments, of an elastic body traveling in the vicinity of a black hole.  Three black hole rotation regimes are shown: for non-rotating black hole ($a=0$, blue), and for quickly rotating black holes with parameter $a \rightarrow - 0.5$ (orange) and $a \rightarrow + 0.5$ (brown).}
\label{TidalDisrupt}
\end{figure}

In conclusion of this section, without listing the details, let us present a plot of the resolved equation of Lagrange, derived 
from Eqs.~(\ref{LagFin}) and (\ref{Lagn}), for the case of destruction of an elastic body traveling near the fast--spinning black hole (Fig.~\ref{TidalDisrupt}).

The body is 
decomposed 
into two fragments of approximately equal masses, $m_1 = 0.53 m_{12}$ and $m_2 = 0.47 m_{12}$. Initially, the fragments were elastically connected with the factor of elasticity $k = 0.1 m_{12}$.
Initially, the body travels along parabolic trajectory with full energy at infinity 
$E = (m_1 + m_2)$ and specific per mass unit angular momentum $j$.
To facilitate visual resolution of the graphs, the calculations used $j = 1.8$.
The more massive fragment is depicted by the larger dot.
Three regimes are presented: a non-rotating black hole ($a=0$, blue curve), and 
quickly rotating black holes, $a \rightarrow \pm 0.5$ (orange curves for retrograde-rotating black hole, $a < 0$; brown curves for prograde-rotating black hole, $a > 0$).
Due to the tidal forces from the black hole, destruction of the elastic system into two detached fragments occurs in all three cases.
But, as Fig.~\ref{TidalDisrupt} shows, the final directions of fragment trajectories depend on the regime of the black hole rotation.

\section{Conclusion}

This paper was envisioned with dual goals in mind: to obtain concrete results and to highlight 
some methodological aspects 
involved in the 
finding of these results.  
The material concerning 
the space-time metric near black holes, coordinate choice, etc., 
contained 
in the 
introductory 
part of the paper, can be found in  works on general relativity, at least implicitly. However, 
these 
foundation elements are scattered (sometimes in implicit form) through various works  (cited in our bibliography and in references therein). We aggregated  this dispersed material. An important purpose of this paper is to illustrate the "Lagrangian" method involved in the conducted study. We wished to convey to the readers who desire fresh perspectives and are unencumbered by ingrained authoritative dogmas, the idea that Method is more important than Result.\footnote{ L. Landau: "A method is more important than a discovery, since the right method will lead to new and even more important discoveries." \url{http://www.azquotes.com/quote/1263664}} 

In this paper, we laid out a step-by-step approach, based on the relativistically invariant, "square--root--form", Lagrangian Eq.~(\ref{i3d}), for studying problems of compact and compound stellar body motion in the vicinity of rotating black holes. We presented a classification of possible regimes of motion, and demonstrated how a body, once in the vicinity of a rotating black hole, becomes involved in the whirlpool--like rotation of the space-time (Fig. \ref{zakhvat}). We determined the conditions for the body's capture and involvement in the co-rotative motion with the black hole. We showed that in describing interaction between components of a multi-body system moving near the black hole, one can use Lagrangian in the form of Eqs.~(\ref{LagFin})--(\ref{Lagn}) derived in the post-Newtonian approximation with respect to parameter $r_g / r$.

Obviously, bodies traveling along curved trajectories must emit gravitational waves (\cite{ll69}, \S 110) and lose energy in the process. This energy loss per unit of time is noticeable only at the fifth order of magnitude in the Lagrangian expansion with respect to small parameter $\sim (v/c)$. With respect to the terms with the first four orders, the energy of the body remains constant. This means that in the expression for the Lagrangian, it suffices to include the terms up to the fourth degree of smallness with respect to $c^{-1}$, and not be burdened with more precise calculations.

For a metric tensor written in the form
$g_{\alpha \beta} = \eta_{\alpha \beta} + h_{\alpha \beta}$  (here $\eta_{\alpha \beta}$ is Minkowski tensor), Lagrangian $L = - m c ((\eta_{\alpha \beta} + h_{\alpha \beta}) u^\alpha u^\beta )^{1/2}$ with $u^\gamma = (1, \dot{q}^i)$. 
Here, the affine parameter $\tau$ is the time measured by the remote observer, i.e. $\tau = t$,  and dot denotes the derivative with respect to this time.  
Obviously,   $\eta_{\alpha \beta} u^\alpha u^\beta = c^2 - v_j v^j$ with $v_j v^j = \textbf{v}^2$, where $\textbf{v}$ is the usual  3-velocity in the 4-coordinate system of the remote observer. Therefore, the Lagrangian for one body in the (given) gravitational field 
takes form
\begin{eqnarray}
\label{fn1}
L = - m c \frac{d s}{d t} = 
- m c^2
\bigg(
1 + h_{00} + 2 h_{0 i} \frac{v^i}{c} - \frac{v^2}{c^2} + h_{ij} \frac{v^i v^j}{c^2}
\bigg)^{1/2} \, .
\end{eqnarray}
Expanding the radical into series up to and including $c^{-4}$, yields: 
\begin{eqnarray}
  \label{fn2}
L = - m c^2 + \frac{m v^2}{2} + \frac{m v^4}{8 c^2} - 
 m c^2 \bigg(
\frac{h_{00}}{2} + h_{0 i} \frac{v^i}{c}  + \frac{1}{2} h_{i j} \frac{v^i v^j}{c^2}
- \frac{h_{00}^2}{8} + \frac{h_{00}}{4} \frac{v^2_a}{c^2}
\bigg) \, .
\end{eqnarray}

This explains the choice of approximation~(\ref{i6g}).

In the context of the presented framework and findings, a question may be asked: what additional information can be extracted from the observational results presented in Fig.~\ref{bhMW}, beyond the conclusion that bright stellar objects orbit around a massive compact object with $M \sim 4.1 \times 10^6 M_\odot$, presumably a black hole?

Let's make a few quantitative estimates.

The Schwarzschild radius, $r_g = 2 G M / c^2$, for a body with mass $M$ is of order $\simeq 2.95 (M / M_\odot) \, km $, i.e., for the black hole in our galaxy, $r_{BH} \simeq 10^7 \, km \simeq 0.07 \, AU$.
The moving objects in Fig.~\ref{bhMW} follow approximately elliptic trajectories.
Only two trajectories are closed, i.e., objects SO-2 and SO-102 have completed their entire rotations over the duration of monitoring. The semi-major axis, $a$, of an elliptic orbit and the distance of the closest approach (periapse), $r_{min}$, with the gravitating center $M$, are related as $a = r_g / u_1 (1 - \epsilon)$, where $\epsilon$ is orbit eccentricity and $u_1 = r_g / r_{min}$.
For a star that does not approach the black hole too closely, simple estimates, following from the classical celestial mechanics, can be used instead of cumbersome formulae of the relativistic theory.\footnote{
In general, however, 
the possibility that 
"dark matter" 
may have an impact 
on the motion of stars  in the central zone of our Galaxy  
should also be noted and considered 
(see, for example, \citet{tp12} and references therein).
}
Then, from the law $a^3 / T^2 = (8 \pi^2)^{-1} r_g c^2$, follows the relationship between characteristic time/period and the distance of the closest approach --  $T^2 (1 - \epsilon)^3 = (8 \pi^2) u_1^{-3}(r_g / c)^2$. Numerically, $T = 9.461 \times 10^{12} \tau $, where  $T$ is measured in seconds and $\tau$ in years. Then we obtain relationship $\tau^2 (1 - \epsilon)^3 u_1^3 = 7.72 \times 10^{-24} (M_h / M_\odot)^2$, which obviously is strictly valid only for the newtonian mechanics, but can be also used for estimations even in post--newtonian models.
This relationship permits not only to estimate $M_h$ based on measured $\tau, \epsilon, u_1$, but also to assess (in terms of the order of magnitude) which values of $\epsilon$ are acceptable for experimentalists, so that the effects (similar to those shown in Fig.~\ref{ProlTraj}) of the general relativity theory become apparent. 
For 
regimes with parameter $u_1 \geq 0.2$ for the observed stars during
$\tau \sim 0.1 \div 1$ (about a few months or a year),
the above-mentioned relationship indicates that the eccentricity of the analyzed orbits must be not small.
To the contrary, parameter $\epsilon$ must be close to $1$, $\epsilon \rightarrow 1$, i.e. trajectories of the observed stars should resemble either very elongated ellipses, or fly-through, quasi-parabolic paths. 
The movement of stars near the black hole  would then occur very quickly, even taking into account the red-shift affecting the distant observer. 
Therefore, high-resolution observations of "bright" moving objects in the region shown in Fig.~\ref{bhMW} must be conducted continuously, not occasionally or with long pauses.   

At present, 
several teams conduct observations of the 
motion of stars orbiting around the supermassive black hole in our Galactic center 
that permit probing 
the gravitational theory (\citet{g14}, \citet{peswhpgdfogg17}, \citet{hd17}, \citet{peskzzs17}, \citet{zly15}).  
In view of their 
estimations (see for example \citet{peskzzs17}) the actually measurable value of parameter $u_1$ (the inverse periapse normalized by the Schwarzschild radius) is of order $7 \times 10^{-4}$ (for object S2 in the S-star cluster in the Galactic center region, which has the shortest period). 
Therefore, 
observation techniques need to evolve 
further
to be able to capture stellar motions with parameters close to the above-mentioned $u_1$ and $\tau$.

\begin{figure}[h!]
\centering
\includegraphics[width=9cm]{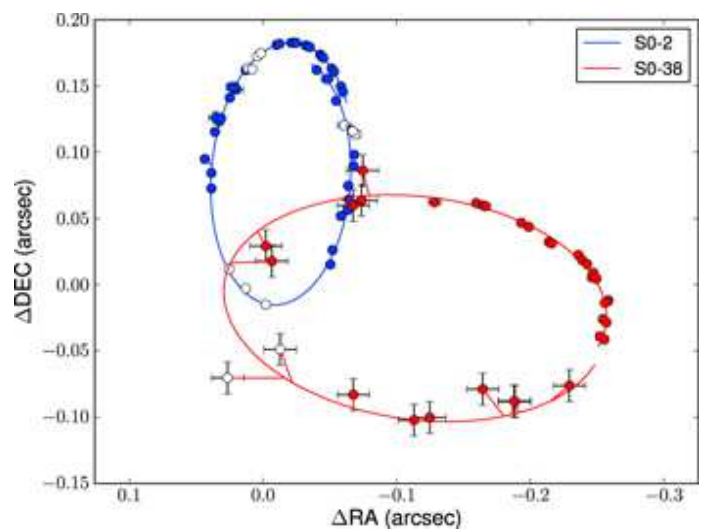}
\caption{Positions on the plane of the sky  and best-fit orbits of objects S0-2 (blue) and for S0-38 (red) for the period from 1995 to 2014.  From\citet{bgsmyambdlmmsw16}.  
}
\label{S2andS38}
\end{figure}

The challenges of obtaining reliable experimental data, which may escaped attention of theoreticians and numerical modelers, can be clearly seen in
Fig.~\ref{S2andS38} 
 (from\citet{bgsmyambdlmmsw16}).  
Fig.~\ref{S2andS38}  shows the positions of the two stellar objects, S0-2 (blue) and for S0-38 (red), from 1995 to 2014. 
The lines show best-fit orbits for the stars. 
Both stars orbit clockwise on the plane of the sky. 
For S0-38 (red), 
some observations show greater uncertainty ranges due to the star's proximity to other objects. 

In general, equations following from Lagrangians~(\ref{LagFin}) and (\ref{Lagn}) can be useful for studying the black hole's tidal effects onto elastic bodies and analyzing the bodies' potential destruction, as illustrated in Fig.~\ref{TidalDisrupt}.

Furthermore, future discoveries of  "bright" objects traveling along \emph{fly-through} trajectories close to this super-massive black hole (similar to those shown in Fig.~\ref{ProlTraj}) would serve as convincing \emph{direct} evidence for the general relativity theory and as an additional confirmation of the existence of black holes.

\vspace{6pt}
\conflictsofinterest{The authors declare that there is no conflict of interests regarding the publication of this article.}

%=====================================
% References, variant A: internal bibliography
%=====================================
\reftitle{References}

%%%%%%%%%%%%%%%%%%%%%%%%%%%%%%%%%%%%%%%%%%
%% optional
%\sampleavailability{Samples of the compounds ...... are available from the authors.}

%%%%%%%%%%%%%%%%%%%%%%%%%%%%%%%%%%%%%%%%%%
\end{document}